\documentclass{aa} 

\usepackage{graphicx}
\usepackage{txfonts}
\usepackage{hyperref}
\hypersetup{
    colorlinks=true,
    breaklinks=true,
    linkcolor=blue,
    citecolor=blue,
    urlcolor=blue,
}
\usepackage{orcidlink}

\newcommand{\cair}{\ion{Ca}{ii}~8542~\AA\xspace}

\newcommand{\cak} {\ion{Ca}{ii}~K\xspace}

\def\specchar#1{{\sc{#1}}}  
\def\specand{\,\&\,}        
\def\Halpha{\mbox{H\hspace{0.1ex}$\alpha$}\xspace}

\def\HeI{\mbox{He\,\specchar{i}}\xspace}

\def\CaIR{\mbox{Ca\,\specchar{ii}\,\,8542\,\AA}\xspace}

\def\CaIIK{\mbox{Ca\,\specchar{ii}\,\,K}\xspace}

\def\MgIIhk{\mbox{Mg\,\specchar{ii}\,\,h{\specand}k}\xspace}

\def\SiIV{\mbox{Si\,\specchar{iv}}\xspace}

\def\CII{\mbox{C\,\specchar{ii}}\xspace}

\def\kms{km~s$^{-1}$}

\begin{document}

\title{Spectral variations within solar flare ribbons}

 \author{A.G.M.\ Pietrow\inst{1}\orcidlink{0000-0002-0484-7634}
 \and M.K.\ Druett\inst{2}\orcidlink{0000-0001-9845-266X} 
 \and V.\ Singh\inst{3}
 \orcidlink{0000-0002-9911-2285}
     }

 \institute{\inst{1}Leibniz-Institut für Astrophysik Potsdam (AIP), An der Sternwarte 16, 14482 Potsdam, Germany\\
 \inst{2}Centre for mathematical Plasma Astrophysics, Department of Mathematics, KU Leuven, Celestijnenlaan 200B, B-3001 Leuven, Belgium\\
 \inst{3} Department of Mathematics, Physics and Electrical Engineering, Northumbria University, Newcastle upon Tyne, NE1 8ST, UK\\
       \email{apietrow@aip.de}\\
      }

\date{Draft: compiled on \today}

\abstract
  {Solar flare ribbons are intense brightenings of primarily chromospheric material that are responsible for a large fraction of the chromospheric emission in solar and stellar flares. We present an on-disc observation of flare ribbon substructures in an X9.3-class flare observed by the Swedish 1-m Solar Telescope.}
  {We aim to identify categories of ribbon substructures seen in the \CaIR, \Halpha, and \CaIIK lines, focusing on their spatial locations and their (spectro-)polarimetric properties. } 
  {We used COlor COllapsed Plotting (COCOPLOT) software to assist in identifying areas of interest.} 
  {We present five categories of spectral profiles within the general body of the flare ribbon: (1) extremely broadened spectral line profiles, where the standard Fabry-Perot interferometer wavelength windows ($\approx 70$ \kms) are not sufficiently wide to allow for a complete analysis of the dynamics and atmospheric conditions. The mechanisms causing this degree of this broadening are not yet clearly understood; (2) long-lived, dense kernels that manifest as more saturated chromospheric line profiles with lower signal in both Stokes parameters. They are interpreted as footpoints of bunched magnetic field loops, whose chromospheric lines form at greater heights than the nearby areas;
  (3) Doppler-shifted leading edges of the flare ribbon in regions that transiently display lower Stokes signals due to the emission dominating at greater heights in the atmosphere; (4) condensed coronal rain overlapping the flare ribbons in the line of sight, producing exceptionally high Doppler shifts near the footpoints; and (5) compact blueshifted areas close to areas with coronal rain down-flows, which are understood to be material that has been thrown up as a result of the down-flowing material impacting the chromosphere. Additionally, a ribbon formation height of about 700 km with respect to penumbral features is estimated using correlating structures on the ribbon and the underlying photosphere.}
  {When selecting areas of the flare ribbon for more general analysis (especially small regions consisting of a few pixels or low-resolution averages), it is important to be aware of the variety of substructures present within a flare ribbon and of the spatial context that can produce these differences. General behaviors across the ribbon should not be inferred from regions that show localized differences. }


  \maketitle
%

\section{Introduction}

\begin{figure*}
        \centering
        \includegraphics[width=1\textwidth, trim=0.cm 0.cm 0cm 0cm, clip]{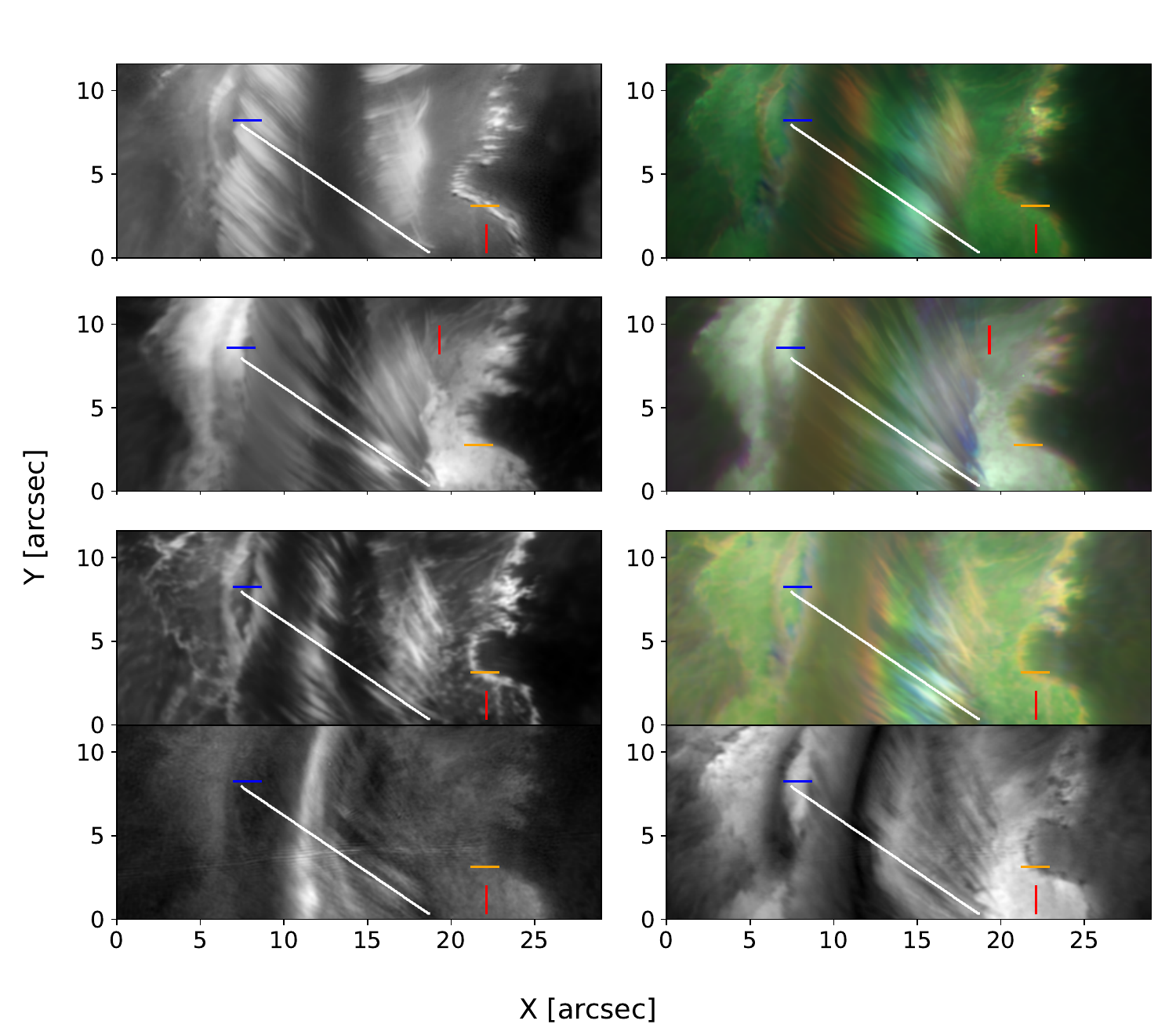}
        \caption{Flare arcade and ribbon {plotted with respect to $(x,y)=515,-268$} between 12:02:52 and 12:02:58 UT in the red wings (left) of the \CaIIK, \Halpha, and \CaIR lines, as well as in a COCOPLOT (right). Colored cross-sections mark representative examples of ribbon freckles (red), the leading edge of the flare ribbon (orange), a blueshifted up-flow region (blue), and the flare arcade (white).  {Top row:} Red wing of \CaIIK along with a COCOPLOT. {Second row:} The same, but for \Halpha. {Third row:} Same but for \CaIR. {Bottom row:} Maps of the \CaIR line of the total linear (left) and circular (right) polarization.}

    \label{fig:overview}
\end{figure*}

\begin{figure*}
        \centering
        \includegraphics[width=1\textwidth, trim=0.cm 0.25cm 0cm 1.cm, clip]{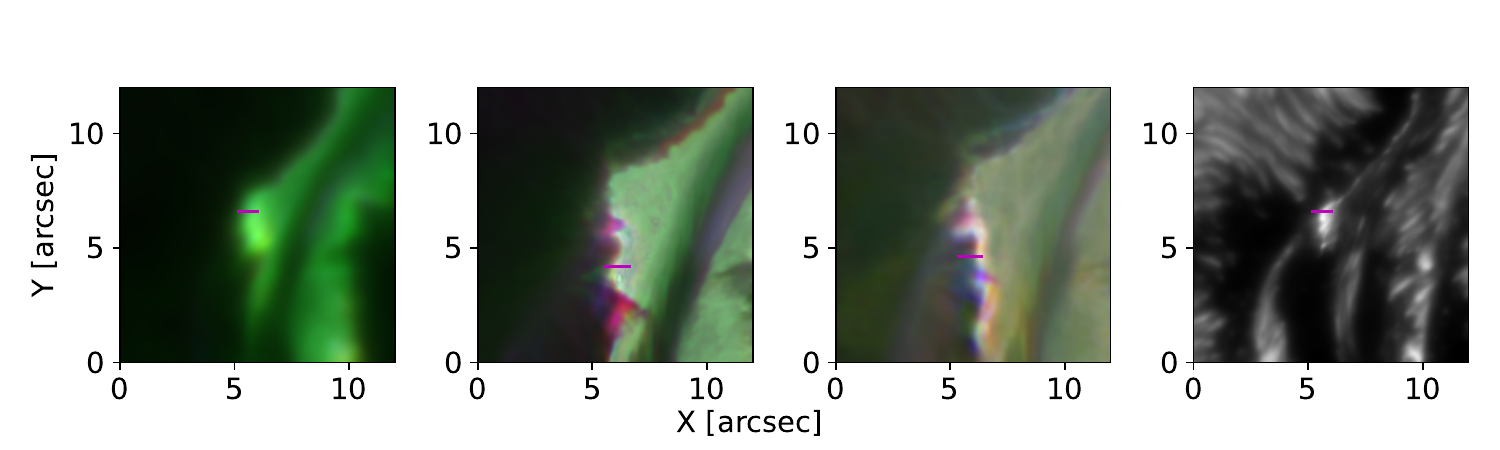}
        \caption{COCOPLOT and 4000~\AA\ continuum images of the flare ribbon {plotted with respect to (x,y)=514,-266} between 11:57:19 and 11:57:25 UT of the \CaIIK, \Halpha, and \CaIR lines. Magenta slits mark the region with the strongest broadening on each image.}
    \label{fig:broads}
\end{figure*}

Solar flare ribbons are regions that exhibit strongly enhanced emission in chromospheric spectral lines during flares. They are of particular interest to researchers studying flares and solar activity as they account for a large fraction of the optical emission and provide a wealth of spectral line data that can be used to diagnose energy transport, conversion, and release. There has been significant recent interest in the use of information from spatially resolved solar flare ribbon observations to understand the physical origins of spectral behavior reported from stellar flares \cite[][]{2017KowalskiBroad, 2019Kowalski, 2022Otsu, Pietrow24harps, Otsu24}. However, there is an energy gap of several magnitudes between solar and stellar flares due to the way that they are detected and measured on stars (as shown in e.g., \citet{Maehara15}, \citet{Pietras2022}, and \citet{Simoes2024}). 

Since the 1950s, researchers have been reporting observations of areas within ribbons that have spectral line profiles exhibiting features such as strong asymmetries \citep{1961Svestka_flare_asymmetries, 1962Svestka_flare_asymmetries, Ichomoto1984, wuelser1989} or especially different profiles of intensity as a function of time \citep{1956Dodson_flares, 1971Harvey_flare_kernels}.
These spectral line behaviors have generally been interpreted as being produced by the lower atmospheric effects (in terms of dynamics, energy deposition, and evolution of atmospheric conditions) resulting from energy released in coronal magnetic reconnection \citep{1958SweetNeutralPoint, 1958SweetParticles, 1963Parker, 1964Petschek} and transported down to the solar surface along magnetic field lines \citep{1966Sturrock, 1968Sturrock, 1971Brown, 1972Syrovatskii, 1978Emslie, 1998Shibata, 2011Fletcher, 2011Zharkova}. 

To date, ribbon formation has been modeled using detailed, field line-aligned, 1D hydrodynamic models with detailed radiative transfer prescriptions and externally prescribed field strength parameterizations, such as RADYN \citep{2005Allred, 2015Allred} and HYDRO2GEN \citep{2017Druett, 2018Druett, 2019Druett}. More recent incarnations of RADYN models have also attempted to bridge the gap to multiple dimensions via stacked 1D modeling \citep{2020Kerr}. Multi-dimensional magnetohydrodynamic studies of ribbon formation \citep{2023DruettAMRVAC, 2024DruettAMRVAC} have recently been produced using the MPI-AMRVAC code \citep{2023KeppensAMRVAC3}, but without the inclusion of the detailed lower atmospheric structure and radiative transfer modeling that is important for an accurate synthesis of energy transport and dynamics in the flaring chromosphere.

Such models provide new insights into the details and remaining questions about flare ribbons, such as their formation height, the velocities and durations of flows within them, and explanations of the exceptionally broad line profiles present within the ribbon \citep{DruettPoster, 2022KowalskiBroad}. This has been achieved in parallel with interpretations and data from space telescopes such as  Hinode \citep{2007KosugiHinode}, the Solar Dynamics Observatory \citep[SDO, ][]{2012PesnellSDO, Lemen2012AIASDO, Scherrer2012HMI}, and the Interface Region Imaging Spectrograph \citep[IRIS, ][]{iris_2014} and ground-based telescopes such as the Swedish 1-m Solar Telescope \citep[SST, ][]{Scharmer03} and GREGOR \citep{Schmidt12, Kleint2020}. 

Flare strengths are typically expressed using the \citet{Baker1970} scale, which is based on the peak soft X-ray flux observed in the 1-8\AA \ channel of the Geostationary Operational Environmental satellite (GOES)  \citep{2005BAMS...86.1079S, 2017BAMS...98..681S, 2019E&SS....6.1730S,Machol2020}. We refer the reader to \citet[][ p29]{Pietrowthesis} for a complete overview of the system.

\citet{2020KuridzeLimbFlare} used SST observations of a flare over the solar limb to determine the height of ribbon emission in the wing of the H$\beta$ line during an X-class solar flare, measuring 300-500~km, in good agreement with previous hard X-ray observations of limb flares \citep{2012MartinezOliverosHXRFlare, 2015Krucker}. It has been known for long that small flare kernels within the ribbon have much stronger red asymmetries in their chromospheric spectral lines, such as hydrogen H$\alpha$ \citep{1971Harvey_flare_kernels}. This has been further highlighted by observations with modern instrumentation \citep{2013DengIBISFlare, 2022OsborneFlareKernels}, which have emphasized such behaviors as the clustering of the profiles with larger red-wing asymmetries in widths of about 1000~km at their leading edges during the impulsive phase \citep{2012Asai}, and the kilogauss strength and visibility of chromospheric magnetic structures within the flare ribbon \citep{2021gragal} between which signs of magnetic reconnection can switch rather abruptly \citep{2012Asai}. \citet{2017Druett} showed that small (sub-arcsecond) individual flare kernels, even in a C1.5 class flare, can exhibit strong Doppler redshifted emission in H$\alpha$ that can be interpreted as downward compressive flows in the chromosphere. 

\citet{2019ZhuMgkhFlareRADYN} and \citet{2019HuangFlareMghkRADYN} studied the \MgIIhk~triplet lines, which show profiles in flares that are particularly difficult to reconcile with self-consistent flare models, with \citet{2019KerrFlareMgIIpartI, 2019KerrFlareMgIIpartII} providing additional insight into modeling approaches that can help to overcome these challenges. \citet{2019ZhuMgkhFlareRADYN} suggest that effects such as the high free electron density and the transient stretching of the transition region (TR) over a greater distance, due to dramatic heating of the chromosphere, can help but not fully explain such emissions. They inferred from their models that the formation region of the spectral line core of \MgIIhk can be compressed and subsequently pushed downwards by around 500~km. A similar conclusion is reached by \citet{Yadav21} for the \CaIIK line. \citet{2019KerrFlareSiIV} investigated the \SiIV line emission with IRIS observations and RADYN modeling, which support the notion that \SiIV is formed under optically thick conditions in stronger flares from plasmas with characteristic temperatures in the range 30-60~kK. 
{Sub-second cadence observations of flare ribbons made with IRIS in \MgIIhk, \SiIV, and \CII line spectra are reported in \citet{2024DruettIRIS}. Interpretation of these observations through MHD models suggests that the combination of TR stretching and the downward motion of the formation regions can contribute to differing TR line Doppler shifts in flare ribbons. Similarly, simulations have shown that the contribution function of photospheric lines can expand well into the chromosphere where peak beam energy deposition occurs \citep{Monson2021}. }

The leading edges of flare ribbons have also been the subject of significant recent study. \citet{2021KerrFlareHeDimming} showed that dimming of the leading edges of expanding flare ribbons in \HeI 10830\AA~can be explained by non-thermal collisional ionization. \citet{2023PolitoRibbonLeadingEdge} found that the leading edges of four flare ribbons showed significantly reduced signs of evaporation and (via their interpretation using RADYN modeling) evidence of reduced electron beam injection effects. \citet{2023DruettAMRVAC,2024DruettAMRVAC} demonstrated that multi-dimensional effects can also impact the leading edge via compression and heat diffusion from neighboring field lines, which should also result in a leading edge with lower electron fluxes and evaporation signatures. \cite{2023_Singh_Riblets} used sub-second cadence optical observations in the H$\alpha$ line to analyze the downward motion of ribbon substructures, which brighten and then move down toward the main body of the flare ribbon over distances on the order of thousands of kilometers at velocities on the order of tens of kilometers per second.

In this paper we combine conclusions from previous investigations with a new analysis of a  X9.3-class flare to present a summary view of solar flare ribbon structures. This interpretation will allow us to (re)interpret previous and newly analyzed data.

\begin{figure*}
        \centering
        \includegraphics[width=\textwidth]{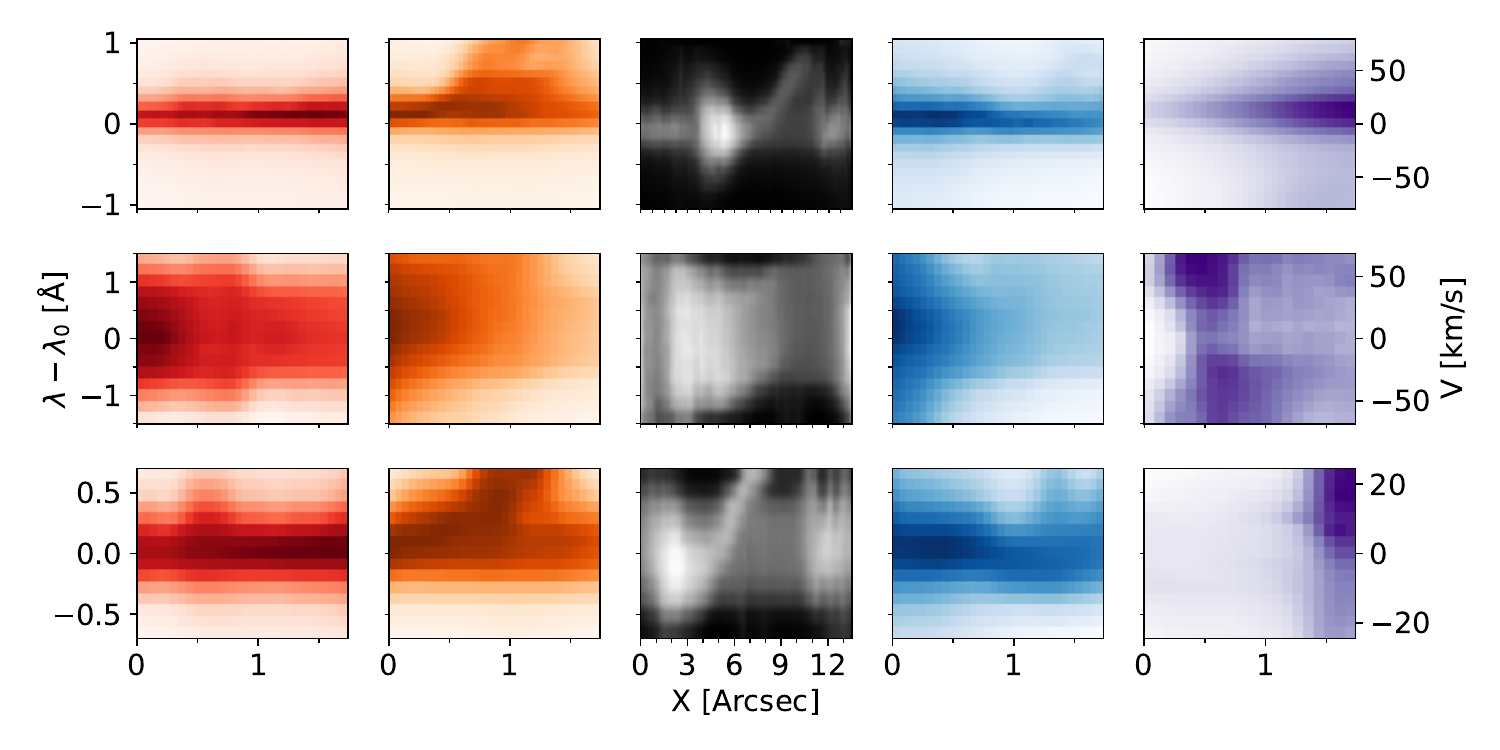}
        \caption{Spectral profiles of the cross-section indicated in Figs. \ref{fig:overview} and \ref{fig:broads}. Each vertical set of three panels corresponds to the \CaIIK, \Halpha, and \CaIR respectively, with each column corresponding to the cross-section of the same color shown in Fig. \ref{fig:overview} or Fig. \ref{fig:broads}. {Left column, in red:} Ribbon freckle at $x\approx0.65$~arcsec. {Second column, in orange:} The leading edge of the flare ribbon at $x\approx0.3$~arcsec. {Third column, in grey:} Cut tracing the trajectory of a coronal loop with condensed rain flowing down towards the flare ribbon. {Fourth column, in blue:} Profiles from the part of the ribbon near the base of a coronal loop down-flow crossing a region showing blueshifted profiles. {Fourth column, in magenta:} Broad profiles found in the early stages of the flare. }
    \label{fig:2dspec}
\end{figure*}

\section{Data and methods}
The active region designated NOAA 12673 ($(x,\, y) = (537\arcsec,\, -222\arcsec)$, $\mu$ = 0.79) was observed on the September 6, 2017, between 11:55 and 12:52 UT with the SST, using the CRisp Imaging SpectroPolarimeter \citep[CRISP,][]{Scharmer08} and the CHROMospheric Imaging Spectrometer \citep[CHROMIS,][]{Scharmer17} instruments.

With CRISP the \Halpha 6562.8\AA~and \CaIR lines were observed sequentially with a total cadence of 15~s. For \Halpha the observing sequence, this consisted of 13 wavelength positions at $\pm$1.50, $\pm$1.0, $\pm$0.80, $\pm$0.60, $\pm$0.30, $\pm$0.15, and 0.00~\AA\ relative to the line center. The \CaIR line was sampled at 11 wavelength positions taken in full Stokes polarimetry mode at $\pm$0.7, $\pm$0.5, $\pm$0.3, $\pm$0.2, $\pm$0.1, and 0.0~\AA\ relative to the line center. The CRISP plate scale is 0.058\arcsec~pixel$^{-1}$, and its spectral resolution is around ${\cal R} \approx 130\,000$.

With CHROMIS only the \CaIIK line, together with the 4000~\AA\  continuum was observed, with a cadence of 6.5~s. The scan consisted of 19 wavelength positions in the \CaIIK line at $\pm$1.00,
$\pm$0.85, $\pm$0.65, $\pm$0.55, $\pm$0.45, $\pm$0.35, $\pm$0.25, $\pm$0.15, $\pm$0.07, and 0.00~\AA\ relative to the line center, plus a single continuum point at 4000~\AA. The CHROMIS plate scale is 0.0375\arcsec~pixel$^{-1}$, and its resolution is around ${\cal R} \approx 120\,000$. 

In addition to the narrow-band images, wide-band images were obtained co-temporally with each CRISP and CHROMIS narrow-band exposure for alignment purposes. The data were processed using the standard SSTRED pipeline \citep{jaime15, mats21} with multi-object multi-frame blind deconvolution \citep[MOMFBD,][]{mats02,vanNoort05}. However, under poor seeing conditions this process fails \citep[e.g., see Fig. 5.4 of ][]{Pietrowthesis}, and for this reason, so-called mixed reductions have been produced according to the methods presented in \citet{Pietrow24harps}. The \Halpha and \CaIR data were first described in \citet{Quinn19}, while the \CaIIK data were first shown in \citep{Pietrow22a} and were reprocessed for \citet{Pietrow24harps}. A detailed review of the data and the evolution of the AR that gave rise to the flare can be found in Section 4.2 of \citet{Pietrow24harps}.

Overviews of the spectral information are given using COlor COllapsed Plots \citep[COCOPLOTs,][]{2022druettcoco}. This method summarizes spectral information by making use of the RGB color space. In our case we have convolved the images with three equally spaced Gaussian wavelength filters, creating three filtergrams centered on the line center and the red and blue line wings. When combined into an RGB image, the relative pixel value of each of these three filtergrams is expressed as a color. For example, a purple pixel represents a typical absorption profile whose wings are above the line center. A strongly redshifted profile will appear red, a nonshifted emission profile will appear green and a heavily broadened profile will appear white since all three filtergrams give a similar value. This makes COCOPLOTs a powerful quick-look tool for identifying and even selecting specific profiles. 

The polarization information is summarized using the total circular  ($\mathrm{TCP}={\int}{|V}_{\lambda }|/{I}_{\lambda } d\lambda$),
and total linear 
($\mathrm{TLP}={\int }\sqrt{{Q}_{\lambda }^{2}+{U}_{\lambda }^{2}}/{I}_{\lambda } d\lambda$) polarizations respectively. {Here, I, Q, U, and V, represent the respective \citet{Stokes1951} parameters. Due to the variable seeing conditions in which this data set was taken, only a limited number of scans exist where all three lines and the continuum point are sampled under sufficient conditions to resolve small scale structures co-temporally. In this work, we focus on two such frames that were taken at 11:57 and 12:02 UT. Four regions of interest were selected in the latter time frame's field of view (FOV), and one in the former. Multiple examples were found of each spectral category, but we focus on giving a representative example of each, showing their spectral and polarimetric behavior to introduce key features of each structure. A future study will focus on the statistical and spectropolarimetric properties of these structure types. 

\begin{figure*}
        \centering
        \includegraphics[width=1\textwidth, trim=0.0cm 0cm 0cm 0cm, clip]{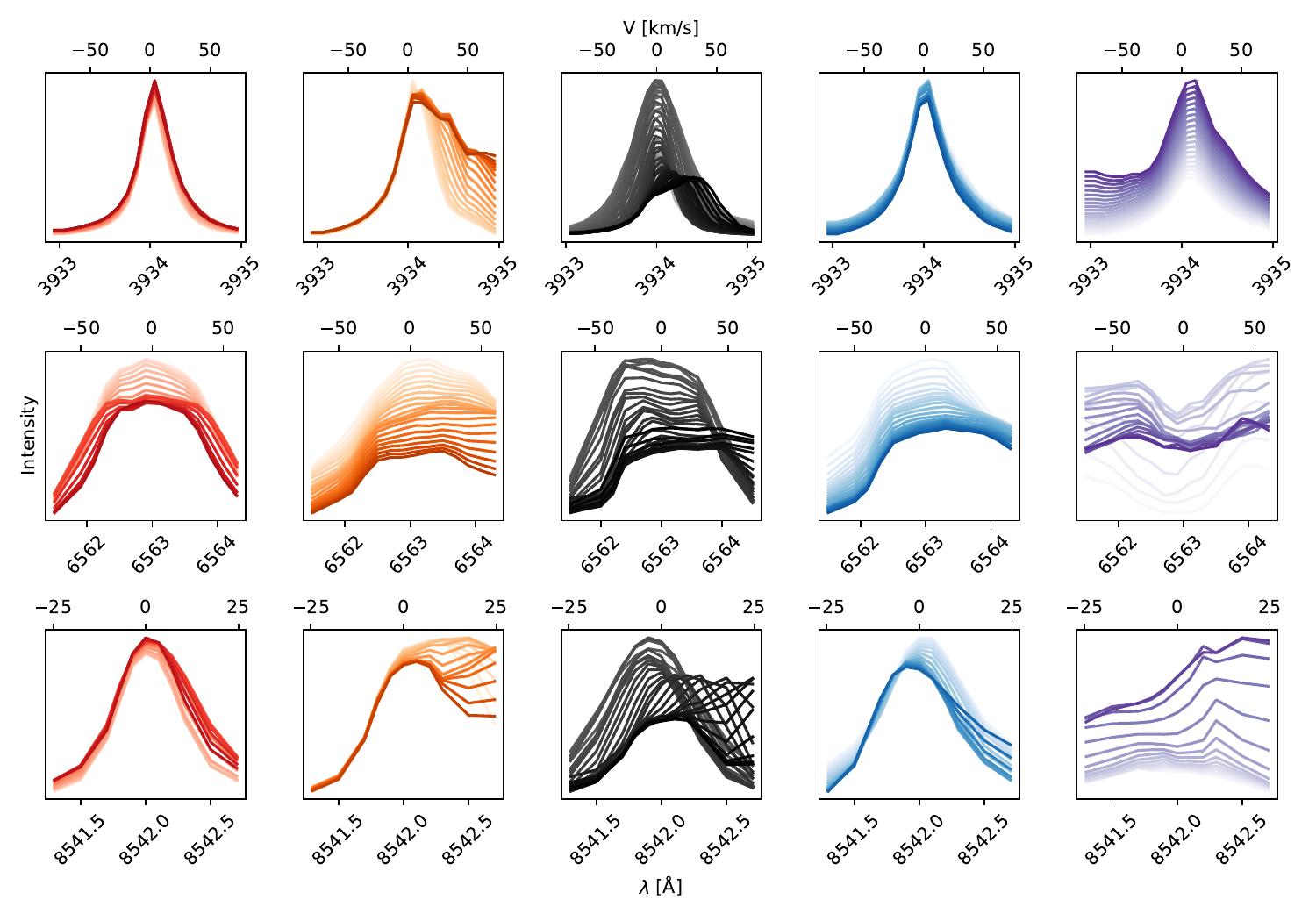}
        \caption{Spectra of each feature described in Fig. \ref{fig:2dspec} compared to their surroundings. The top row shows the \CaIIK spectral lines, the middle row shows \Halpha, and the bottom row shows \CaIR. The left column shows in red these spectra for the red slit from Fig. \ref{fig:overview} placed over a ribbon freckle. The second column shows in orange the results for the leading edge of a flare ribbon. The third column shows in gray the spectra for the slit tracing the trajectory of a flare loop, which shows coronal down-flows over a ribbon area in the line of sight. The fourth column shows spectra in blue for profiles at the base of the coronal rain loops. The fifth column shows broad profiles. The darkest line in each of the panels is the spectral line shape from the pixel in the middle of each structure selected for examination, and the lighter lines represent the spectra as one moves out from this feature along the slit indicated in Fig. \ref{fig:2dspec}.}
    \label{fig:1dspec}
\end{figure*}

\begin{figure*}
        \centering
        \includegraphics[width=1\textwidth, trim=0.0cm 0cm 0cm 0cm, clip]{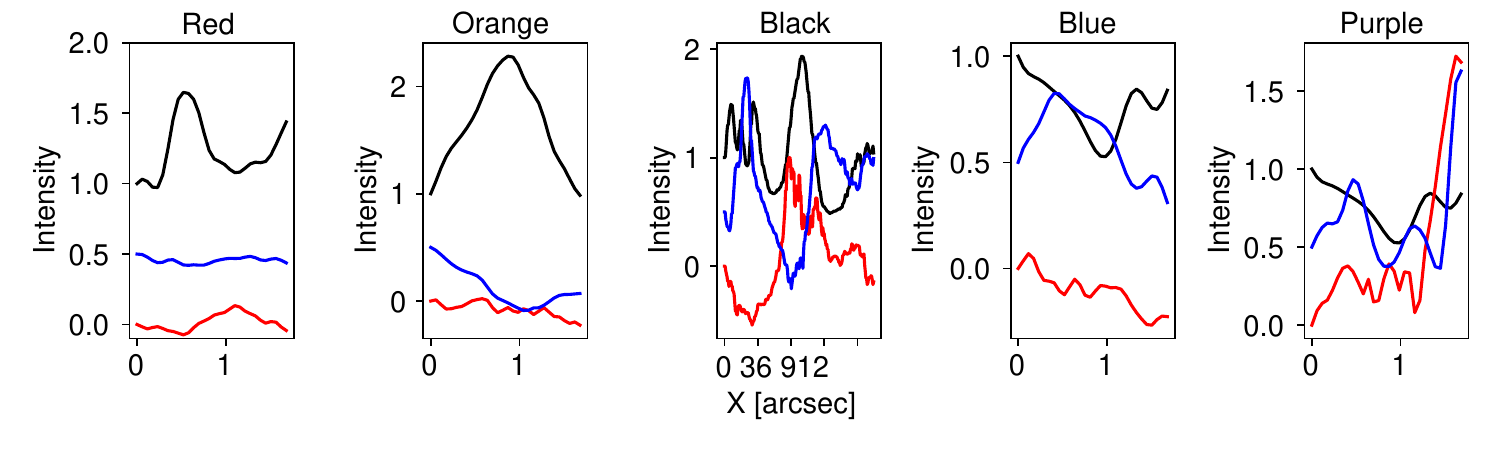}
        \caption{\CaIR polarization degree derived from the Stokes profiles extracted from the pixels along the colored cross-sections shown in Figs. \ref{fig:overview} and \ref{fig:broads}. The black line represents the red-wing intensity along each slit, the blue line represents the TCP, and the red line denotes the TLP along the same cut. Each line profile is normalized and offsets are given for ease of comparison.}
    \label{fig:specpol}
\end{figure*}

\begin{figure*}
        \centering
        \includegraphics[width=1\textwidth, trim=0.0cm 0cm 0cm 0cm, clip]{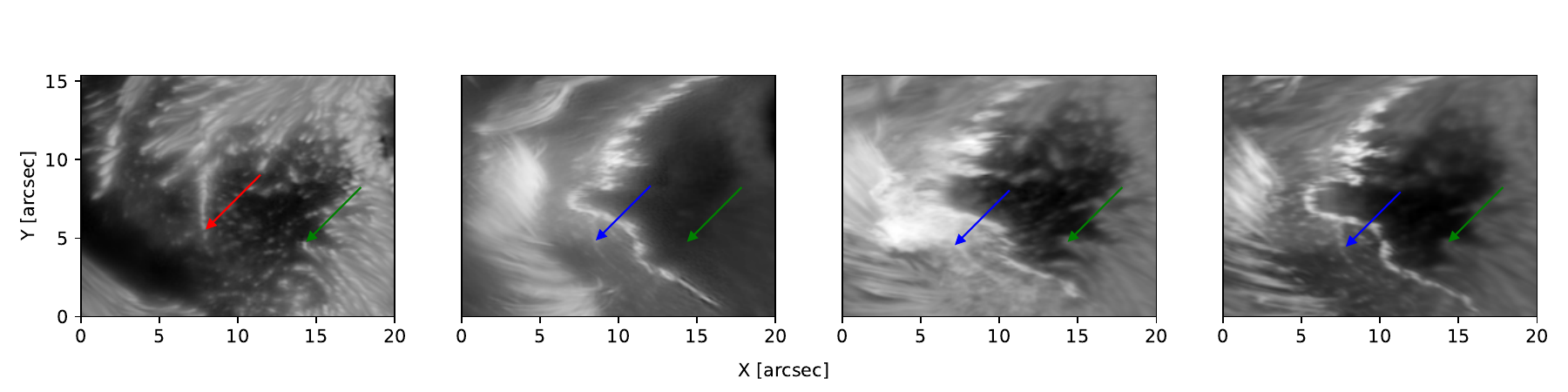}
        \caption{Photospheric map { at 11:57~UT, plotted with respect to (x,y)=529,-278} taken in the 4000~\AA\ continuum (left panel), along with co-aligned, logarithmically scaled maps of the red wing of \CaIIK, \Halpha, and \CaIR, respectively. A chain of peripheral umbral dots is indicated with the red arrow, while its imprints on the ribbon are indicated with the blue arrows. Green arrows indicate the location of a penumbral feature used to align all images. The offset between the red and blue arrows gives an indication of the height difference between each layer.}
    \label{fig:height}
\end{figure*}

\section{Results} \label{sec:results}

In Fig. \ref{fig:overview}, an approximately 30 by 12 arcsec FOV centered on the flare ribbons is shown in \CaIIK, \Halpha, and \CaIR (top three rows) with total circular and linear polarization in the \CaIR line shown in the bottom row. {Each FOV is given with coordinates relative to (x,y) = 515,-268.} A red wing intensity image and a COCOPLOT are shown to distinguish between the different ribbon sub-structures, and the total polarization maps indicate the magnetic field structure. The spectral lines in this figure are sorted by typical relative solar formation heights (although actual formation heights can vary greatly within the 3D structures of the chromosphere). The \CaIR line core forms in the lower chromosphere, below \Halpha. The \CaIIK line core forms above these two, in the middle of the chromospheric layer \citep[Fig. 1]{Jaime19}. While the  modeling \citep{2005Allred, 2015Allred, 2017Kowalskidown, 2017Simoes, 2017Druett, 2018Druett, 2019Druett, 2024DruettAMRVAC} and the mean heights of flare ribbon emission observed over the solar limb \citep{2020KuridzeLimbFlare} do suggest that the lower atmosphere of the flare is greatly compressed when compared to the quiet Sun, we still assume for the sake for this graphic that the relative ordering of the line core formation regions remains intact. 

Five structures of interest are investigated in this work by extracting spectral profiles from the lines of pixels intersecting the feature locations (see Fig.\ref{fig:overview}). {While most slits overlap for all spectral lines, we found that it was not possible to achieve this for the first type of region (indicated in red). A different location was picked for the \Halpha line to ensure that the selected profiles are representative. }
\begin{enumerate}
    \item Dense kernels that create a variegation pattern on the bright flare ribbon (indicated by red lines). We name these regions "ribbon freckles" due to their spotty nature on the otherwise smooth flare ribbon. They can be seen as bright spots in the \CaIR red wing, and as purple (\Halpha) or yellow (\cair) spots in the COCOPLOTS. {These features are much less obvious in both the wing and COCOPLOT of \CaIIK, but they can be seen as a subtle brightening in the former.  }

    \item The leading edge of the ribbon, which exhibits spectral profiles with larger red-shifts than the typical profiles on the main body of the ribbon, is indicated with orange lines. Therefore, this feature appears as a bright edge in the red wing of all three lines and with an orange-red color in the COCOPLOTs. 
 
    \item Areas with spectra that exhibit large redshifted components in their profiles which occur near the footpoints of the flare loop arcade (profiles taken along the white line). 
    In the COCOPLOT, they can be seen as a color similar to that of the freckles, making it more difficult to distinguish between the two, close to the flare arcade. 

    \item A narrow lane of Doppler blueshifted profiles found on either side of the flare arcade, close to the down-flows of region three (indicated with blue lines). 

    \item Wide profiles that occur in small bright flare kernels early on in the time series. These are indicated with magenta lines in Fig. \ref{fig:broads}, where they have a white color in the \Halpha and \CaIR COCOPLOT, while appearing up as a bright green color in \CaIIK. In the continuum, they correspond to white light flare emission. 
\end{enumerate}

The spectra of the features shown with cross-section cuts in Figs \ref{fig:overview} and \ref{fig:broads} are displayed in Figs. \ref{fig:2dspec} and \ref{fig:1dspec}, where each column corresponds to the aforementioned features and each row shows the \CaIIK, \Halpha, and \CaIR spectra. 

\subsection{Ribbon freckles and flare ribbon height}\label{sec:freckle}
The red lines in Fig. \ref{fig:overview} correspond to ribbon freckles. The colored slits cross a freckle at roughly X=0.6-0.7 arcsec, where there is some broadening in each of the three lines. To illustrate this, the individual spectra of the 15 pixels before the core of the freckle have been plotted in the first column of Fig. \ref{fig:1dspec}, where the darkest profile represents the center of a freckle, and the lightest profile represents the background area slightly away from it. From this, it can be seen that in both calcium lines the freckle is characterized by a broadening in the red wing, while in \Halpha it broadens on both sides, as well as flattens off in the core. Visually, there is no clear shift in the central position of the \Halpha line. However, when comparing the line profile centroids a close agreement is found for the Doppler shift in each of the three lines, with \CaIIK, \Halpha, and \CaIR showing maximum displacements of 1.55\kms, 1.69\kms, and 1.70\kms respectively with respect to the centroid positions outside the freckle. {The line centroids were calculated using equal weightings for each measurement. Each measurement is the average of the locations of the left and right sides of the line profile at intensities equal to (0.5,0.6,0.7,0.8, and 0.9) of the maximum value. These fractions are measured in terms of the increase above the baseline intensity, which was taken in far the blue wing. These threshold values were chosen so that the profile locations we used may all fall inside the available spectral window, while the linear interpolation of the profile values was used to estimate the locations of the intersections of the line profiles, with threshold values between sampled wavelengths.}

From the polarimetric signal, which is plotted from profiles taken along the slits in Fig. \ref{fig:specpol} and the maps in Fig. \ref{fig:overview}, it can be seen that these structures tend to have an overall lower signal in both the linear and circular polarization maps relative to their surroundings. When compared to the 4000~\AA~continuum, some but not all freckle structures can be seen to correspond to the locations above visible umbral dots and penumbra structures. This is illustrated in Fig. \ref{fig:height}, where the imprint of a narrow chain of peripheral umbral dots can be seen on the flare ribbon in all three lines. 

The flare ribbon height can be estimated based on the location of ribbon freckles if it is assumed that the freckle is oriented vertically and that the ribbon lies parallel to the solar surface. This is done by equating the viewing angle to the hypotenuse of the shift in the FOV $\Delta r$, and the height between the two objects $\Delta z$, {with a consistency check performed to ensure that the quotient of these changes is approximately in agreement with what would be expected for a vertical structure at this viewing angle $\theta$. Solving this right-angled triangle gives $\Delta z = \Delta r / \sqrt{ 1- \cos^2\theta}$.  }

For \CaIIK, a precise shift of 0.75" can be estimated with respect to the corresponding photospheric feature, as both the line and its continuum are observed in the same scan. For the other two lines this is more difficult due to the absence of a continuum point to align with, as well as the differing pixel size and sub-optimal seeing conditions between them and the 4000~\AA~ continuum point. Instead, {the peak intensity of} a familiar photospheric point (Fig. \ref{fig:height}, green arrow) is selected in the line wing of all three lines, while the distance between that and the imprint of the chain of peripheral umbral dots is calculated. The three measured horizontal shifts for \CaIIK, \Halpha, and \CaIR are 90, 91, and 93 CRISP pixels respectively, which correspond to heights of 710, 690, and 600 km above the photospheric reference point. 

In \CaIIK, the feature marked by the green arrow moves by roughly 0.2" {(5 CRISP pixels)}, when observed from the continuum point versus the line wing {due to the change in formation height of the continuum and the spectral line.} {Conservatively, a similar shift is assumed in the remaining two lower-forming, lines, which is combined with the uncertainty of finding the exact location of the penumbral feature. } {We use this value, which corresponds to roughly 150~km, as an uncertainty in the measures.} For the \CaIIK line, a smaller uncertainty of 3 CHROMIS pixels is assumed, which corresponds to roughly 100~km.

\subsection{Leading edges}
The orange line in Fig. \ref{fig:overview} corresponds to a representative location for the leading edges of the flare ribbon, with the corresponding spectra being shown in the second column of Figs. \ref{fig:2dspec} and \ref{fig:1dspec}. In all three lines, a strong enhancement is seen in the red wing that is not observed in the blue wing or central part of the profile, suggesting additional components to the emission spectrum. The separation of these components is particularly visible in the \CaIR line (Fig. \ref{fig:2dspec} central column, bottom pair of panels) and somewhat visible in the \CaIIK line (Fig. \ref{fig:2dspec} central column, bottom pair of panels). In the polarization, we see a strong dip in total circular polarization in this area, while the total linear polarization does not change as drastically.

\subsection{Down-flows}
The white cross-section in Fig. \ref{fig:overview} is aligned with the post-flare loop arcade, which has footpoints overlapping the trailing edges of the flare ribbons. It is longer than the other cuts to capture the full behavior along the condensed material in the loops. The cut starts on the right side, which is represented with light grey profiles in the third column of Figs. \ref{fig:2dspec} and \ref{fig:1dspec}. Both representations show profiles with central enhancement along the full length of the slit, with a component that shows a strong  red-shift towards the end of the cross-section at the base of the loop footpoint. The red-shift of this enhanced emission component increases until the signal exits the observational spectral window in both the \CaIR and \Halpha lines, but stops at around 70 \kms in \cak. In polarization, it is found that the signal is primarily linearly polarized in the first part of the cross-section, near the coronal loop-tops of the arcade. The signal switches to being primarily circularly polarized as we move along the slit towards the footpoint of the loop and the trailing edge of the flare ribbon, as one would expect for a signal from material following a magnetic field loop through the corona viewed at this orientation.

\begin{figure}
    \centering
    \includegraphics[width=\columnwidth, trim=9.5cm 0cm 7.3cm 0cm, clip]{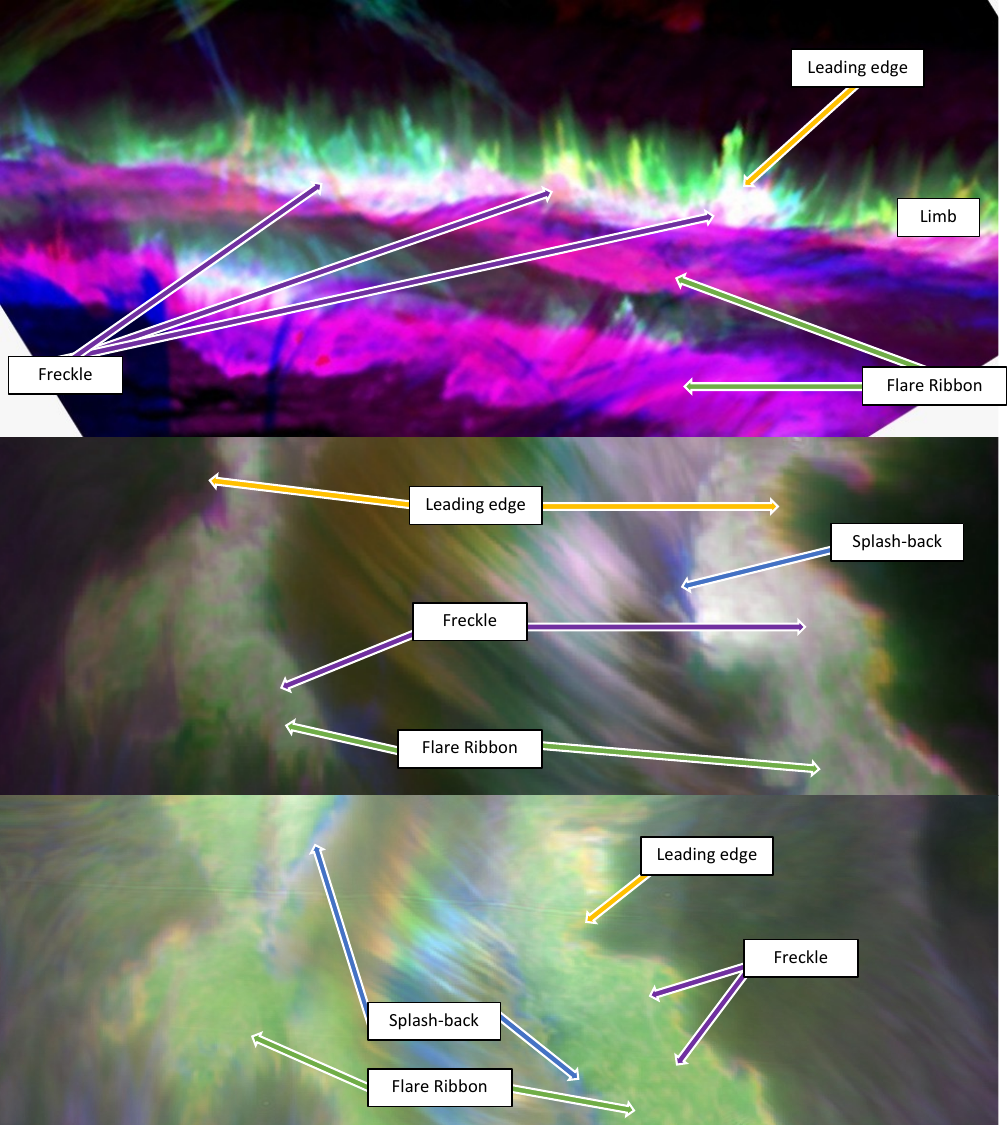}
    \caption{Comparison between two COCOPLOTS of X-class flares in edge-on, and top-down views with the described region types pointed out from a top-down and sideways view. {Top:} Two-ribbon X1.5-class flare seen in \Halpha at the solar limb in the NOAA Active Region NOAA 12087 observed with CRISP at the SST on June 10 2014 at 12:52 UT. The data for this flare is comprehensively described in \citet{2023_Singh_Riblets}. {Middle: }\Halpha observations of our X9.3-class flare. {Bottom:} \CaIR observations of our X9.3-class flare}
    \label{fig:limbflare}
\end{figure}

\subsection{Confined up-flows}
The blue line in Fig. \ref{fig:overview} crosses the narrow lane showing blueshifted emission profiles in \CaIR that is located close to the footpoints of the coronal rain downflows. The measured blue shift is on  a similar order as the freckle red-shift and it is stronger in \CaIR than \CaIIK. In \Halpha, it is not measured, and only a red-shift is visible. The polarimetric signal stays mostly the same when compared to its surroundings.

\subsection{Broad profiles}
The magenta cuts in Fig. \ref{fig:broads} cross bright kernels close to the flare ribbon edge that are white in both the COCOPLOT images of \CaIR, and \Halpha, and a bright green in \CaIIK. The profiles taken from these areas are shown in Figs. \ref{fig:2dspec} and \ref{fig:1dspec}, which corroborate that this coloration is due to a strong line profile broadening in these locations.

\section{Interpretation, discussion, and conclusions}


\begin{figure}
        \centering
        \includegraphics[width=1\columnwidth, trim=5cm 2.5cm 6cm 1cm, clip]{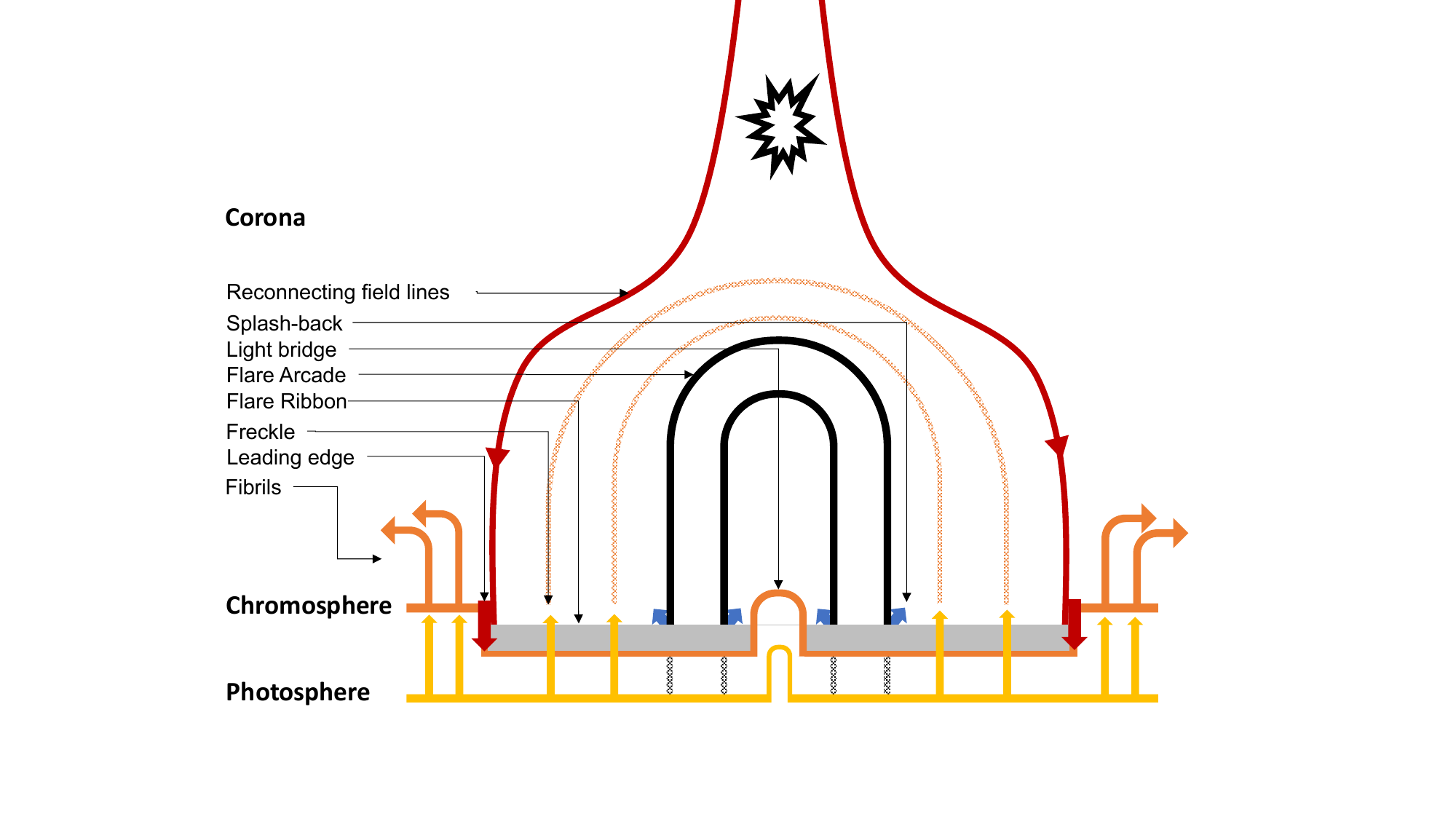}
        \caption{Schematic representation of a cross-section through a two-ribbon flare, with the features described in this work. }
    \label{fig:cartoon}
\end{figure}

The high spatial and spectral resolutions of these observations allow for a detailed study of flare ribbon substructures at various heights in the chromosphere through a set of three spectral lines. This on-disc view is also contrasted with an on-limb X1.5-class flare observation from \citep{2023_Singh_Riblets}, which can be used for further validation of our suggested model for their origins. These two points of view, together with characteristic examples of each of the five features described in this paper, are shown in Fig. \ref{fig:limbflare}. Additionally, a schematic cross-section of a two-ribbon flare presenting our findings regarding these substructures is shown in Fig.~\ref{fig:cartoon}. In this section we break-down the discussion of these features, separating each into its own subsection, starting with the five types described in the results (Section \ref{sec:results}) and ending with a discussion on the height of the flare ribbon emission. 

\subsection{Ribbon freckles}
An example of these ribbon sub-structures is marked in red in Fig. \ref{fig:overview}, with their profiles displayed in  the first column of Figs. \ref{fig:2dspec}, \ref{fig:1dspec}, and \ref{fig:specpol}. In the schematic diagram shown in  Fig. \ref{fig:cartoon}, they are represented by the tops of the yellow bunched field on the flare ribbon, which matches the elevated purple structures seen in the on-limb observation of Fig. \ref{fig:limbflare}.
In \Halpha, the most noticeable characteristics are the flattening (saturation) and broadening of the line core. {This can be interpreted as an increased opacity due to the higher density of material in this region, which, in turn, leads to a saturation of the profile against the source function \citep[See ][ Fig. 7]{Pietrow22a}.}
Additionally, a small centroid red-shift of 1.69 \kms was also measured. Some of these regions align with umbral and penumbral structures found in the photosphere. 
In both the \CaIR and \CaIIK lines, the line core intensity increases slightly and in all three spectral lines the line broadens on the red side of the profile, suggesting the presence of low-velocity down-flow components, or velocity gradients in these regions. Maximum displacements of 1.55 \kms, 1.69 \kms, and 1.70 \kms were found for the \CaIR, \Halpha, and \CaIIK lines respectively with respect to the centroid positions outside the freckle.
{These results, coupled with the decrease of both the linear and circular polarisation inside the freckle as compared with the rest of the slit (as seen in Fig. \ref{fig:specpol}) implies that the plasma in these regions is less magnetized than its surroundings. }
We propose that these features occur above concentrations of flux on the ribbons, where a greater density of material is collected during ribbon formation, compared to the surroundings.  This is consistent with the \Halpha line opacity, which is strongly correlated with column mass \citep{jorrit12}. This means that such a flux tube would raise the opacity locally and, in turn, broaden and saturate the line core via opacity broadening. A larger opacity contribution from the upper regions of the flare ribbon can also explain the lower polarization signal despite the flux concentration below. This is because the formation region is sampled higher in the atmosphere and the emission is thus forming at heights with a lower magnetic field strength. This is also consistent with features we have observed in the limb flare data set presented by \citet{2023_Singh_Riblets}, where it is possible to see ribbon freckles from the side. On the solar limb, they are visible as small "puffs" and "loops" of higher density chromospheric material extending above the main flare ribbon structure. 

\subsection{Leading edges}
The leading edges, marked by orange lines in Fig. \ref{fig:overview}, exhibit increased emission along with a strong Doppler redshifted component compared to the rest of the ribbon. Additionally, the circular polarization signal decreases strongly, such as the freckles (see column 2 of Figs. \ref{fig:2dspec}, \ref{fig:1dspec},  and \ref{fig:specpol}). They are represented as red downwards-pointing arrows in Fig. \ref{fig:cartoon}.

In \citet{2023_Singh_Riblets}, it was found that the flare ribbon exhibits rapid brightenings of higher-lying chromospheric material that subsequently decreases in height with velocities on the order of tens of km~s$^{-1}$. In particular, the authors consider that this apparent velocity could be due to plasma motions or thermal conduction including heating or cooling. Our top-down observations provide leading-edge features with Doppler-shifted velocities that are on the same order as those registered by \citet{2023_Singh_Riblets}  for plane-of-sky plasma motion velocities{ (typically several tens of \kms}). This interpretation of a higher formation height is again consistent with the observed lower polarimetric signal. Thus, in our schematic, these are interpreted as features at the leading edges of the flare ribbons. In particular, the compression of the non-flaring chromospheric density plasma and pre-existing chromospheric structures at the leading edges are displayed using red arrows near the flare ribbon in Fig. \ref{fig:limbflare}. These have traditionally also been interpreted as sites of energetic particle beams, although more recently other potential effects at the leading edge have been presented \citep{2023PolitoRibbonLeadingEdge, 2023DruettAMRVAC, 2024DruettAMRVAC}. Moreover, the side-on view provided by the limb flare in the top panel of Fig. \ref{fig:limbflare} is consistent with the impression of greater formation heights that one can also gain from the on-disc view; namely, the chromospheric structures outside of the flare ribbon area form at greater vertical heights in the atmosphere. This can be seen via the broad flare ribbon emission that appears white in the COCOPLOT. However, the core wavelengths of this emission are absorbed by the cooler material in front of it in the line of sight, which (due to the viewing angle) must be related to the plasma at higher vertical locations. This absorption removes the signal from the COCOPLOT central (green) filter, leaving areas with overlapping cool material in the line of sight as bright magenta in the image, whereas the unobstructed areas of the flare ribbon are bright white.

\subsection{Coronal rain down-flow regions}

Close to the footpoints of the flare arcade there are chromospheric spectral profiles that indicate the presence of strongly redshifted {components of the emission. The maximum measurable Doppler velocities within the spectral windows are around 70 \kms and contributions from the emission in \Halpha and \CaIR appear to be present outside the limits of their respective wavelength windows. T}hese are marked by the white line in Fig. \ref{fig:overview} and their spectra are shown in the third column of Figs. \ref{fig:2dspec} and \ref{fig:1dspec}.

These features are evidenced by our high-resolution observations as coronal rain down-flows from the now-cooling and condensing material that had previously been evaporated into the reconnected flare loops. They are represented as the footpoints of the black loops near the trailing edge of the flare ribbon in the schematic in Fig. \ref{fig:cartoon}.

Typical values for chromospheric down-flows observed with modern instruments are between 40 and 100~\kms \citep{2017Druett,Bart21,Kerr2022}, often with lifetimes on the order of tens of seconds to a minute. However, older \Halpha observations with wider spectral windows showed flare ribbon spectra close to the polarity inversion line with Doppler-shifted components between 100 and 200~\kms, lasting for timescales that are on the order of a few minutes  \citep{Ichomoto1984, zarro1988, wuelser1989, 1989Zarro}.

The older results can now potentially be re-interpreted as regions at the footpoints of coronal rain loops that are overlapping with the flare ribbon in the line of sight. Spectra from our study in these regions show emission enhancement components that move out of the spectral window as we move along the flare loop, toward the foot points of the arcade. This implies that down-flow velocities above 100 \kms are plausible in these regions. Moreover, the prolonged supply of coronal rain mass over a few minutes could also explain the long duration of the redshifted components that were previously observed, without any recourse to an exceptionally deep ingress of the down-flowing material into the Sun. 

In UV lines, velocities above 100 \kms have recently been reported in IRIS observations by \citet{Xu2023} for a short-lived region of interest inside of an X1.3 flare. In that study, these down-flow velocities were also interpreted as decelerating coronal down-flows. 

\subsection{Down-flow splashback}
The narrow channel of areas showing blueshifted emission at the bases of coronal rain down-flows were investigated using profiles along the cuts indicated by blue lines in Fig. \ref{fig:overview}. These profiles exhibit a low-velocity blue-shift in the calcium lines, while no conclusive shifts are seen in \Halpha (see column 4 in Figs. \ref{fig:2dspec}, \ref{fig:1dspec}, and \ref{fig:specpol}). They are represented as blue upwards pointing arrows in Fig. \ref{fig:cartoon}.  
The small area in which these up-flows occur, and their proximity to the coronal rain down-flows suggests that this is not the same type of up-flow described in \citet{Schmieder1990} and interpreted as a gentle evaporation (based on   the transition-region line), whereas this feature can barely be detected in the lowest forming chromospheric line. 

Instead, we propose that the proximity of these regions to the redshifted profiles caused by coronal rain, as well as the parabolic spatial variation of the blue-shifted \CaIR feature, suggest that these may be a form of reflection (or "splashback") from the impact of the coronal rain against the flare ribbon. The fact that the polarization signal is comparable to the surrounding area also suggests that this material does not travel as high as the freckles. Due to the sideways projection, and except where the line-of-sight is favorable, it is not always easy to distinguish these features from the leading edge of the limb flare ribbons in Fig. \ref{fig:limbflare}. This phenomenon could potentially be modeled in future studies using the latest multi-dimensional flare simulations \citep{2020Ruan, 2023DruettAMRVAC, 2024DruettAMRVAC}. 

\subsection{Broad profiles}
Highly broadened spectral lines are common in chromospheric lines, particularly in \Halpha, however,  due in part to their intractable nature, they have not been commonly studied \citep{DruettPoster, 2022KowalskiBroad}. We note that example plots from the same data set were provided in Figure 14 of \citet{2020Zharkov} for \Halpha and additional examples for \CaIR have been reproduced in this work. 

The "appearance" of line profile broadening can arise for a number of reasons. This effect can be created by multi-velocity plasma in the line of sight, such as Doppler-shifted emission from coronal rain or prominence material located over centrally enhanced emission from below (see broad profiles in column 3 of Fig.~\ref{fig:1dspec}), or spatially unresolved flows within an observational "pixel". These should however, be carefully distinguished from flattened profiles without broadening, which can also be produced in lines that are typically absorption lines, but that are nearing conversion into emission \citep[e.g., ][ Fig. 5.]{Berlicki05}. Also, opacity broadening can produce significant variations in line profile widths, without any recourse to highly increased temperatures to explain the variations that would be required using thermal broadening \citep{molnar2019, Pietrow22a}.

However, many examples do appear to be highly "broadened" in the true sense of the word, particularly in the impulsive phases of solar flares. Typically broadening mechanisms considered in spectral lines include Lorentzian, such as natural broadening, and Gaussian effects, such as thermal motions \citep[see Section 3.3 of][for a more comprehensive discussion of these broadening effects]{2003Rutten}. However, the typically considered line-broadening processes are far from able to explain the full widths of such profiles, which often also defy capture in Fabry-Perot interferometer spectral windows.

No comprehensive explanation of these line widths currently exists. Attempts have been made to see if they are consistent with generation through mechanisms such as the Orrall-Zirker effect \citep{1976OrrallZirker}, namely, the exchange of an electron with a proton in a beam of energetic ions accelerated as well as the electrons in the corona \citep{2023KerrOrrallZirker}. Another important factor is the modeling of broadening due to the quadratic Stark effect, due to the presence of high densities of free electrons in the flare ribbon \citep{1993Zharkova, 2016KowalskiStark, 2017KowalskiBroad, 2018Druett}. Investigations are ongoing to address this issue \citep{2022KowalskiBroad, 2023KerrWidths}. In this paper, we also mention a speculative thermal, but particle process that might be considered as a candidate to help explain these excessive line widths. The proposed mechanism is a "companion" of the Orrall Zirker effect, namely, it is a "memory effect" of high-temperature protons, and other ions. The tops of flare chromospheres have a compact interface between chromospheric temperature material and extraordinarily dense and high temperature thermal coronal material that is often highly turbulent

Although electrons at coronal temperatures tend not to recombine with hydrogen ions (protons) due to their great microscopic energies and the slow recombination rates \citep{2007Leenaarts}, high-temperature thermal coronal protons intruding into cooler layers could come into contact with electrons thermalized to chromospheric temperatures that have similar microscopic velocities, due to the $1836$ mass ratio between the proton and electron particle species. It is feasible that these populations could recombine. If recombination occurs between these two populations then exceptionally broadened emission would be produced, with memory-effect "thermal" Doppler widths in great excess of the local electron temperature. However, this effect is yet to be tested or proven with modeling and experimentation. 

\subsection{Flare ribbon height}
The height of the flare ribbon can be estimated for all three lines based on the locations of ribbon freckles and the corresponding structures below (see Fig. \ref{fig:height}.) However, as the photospheric features seen in the line wing of these lines do not form at the same location, an uncertainty of around 150 km ({See end of Sec. \ref{sec:freckle})} is introduced to the height that has been measured in this way. This is not the case for the \CaIIK line, as it is co-aligned with a continuum point that allows for a direct comparison between the two, resulting in a lower uncertainty of roughly 100 km. Heights of 710, 690, and 600 km have been measured for \CaIIK, \Halpha, and \CaIR, respectively, {in relation to the umbral feature marked in Fig. \ref{fig:height}.} This is consistent with the fact that the freckles used to measure these heights have very similar centroid redshifts, which means that they either have a consistent velocity structure over height or they are sampled at similar heights. 

While the error bars make it impossible to confirm whether or not the lines indeed form in the expected order, it is apparent that the chromosphere is indeed compressed down greatly when compared to a non-flaring atmosphere \citep[e.g.,][]{vernazza76, Jaime19, roberta2020}. It is also in line with the ~170km wide flare layer width described by \citet{Berlicki2008}, although our flare ribbons appear to form much lower than is suggested in that work. 

This is higher than the 300-500~km H$_\beta$, and 0~km \CaIR ribbon heights inferred by \citep[][]{2020KuridzeLimbFlare} and from x-ray signatures \citep[200-300~km][]{2012MartinezOliverosHXRFlare}, but slightly below the value of 800~km above the photosphere inferred by \cite{2015Krucker} from the X-ray emission. However, it should be noted that our height estimate is with respect to features inside of a sunspot which are effected by Wilson depression \citep{wilson1774observations}. This is around 600~km (on average) for the umbra \citep{Loptien2018}, but it is expected to be less for umbral dots \citep[e.g., ][]{1969Wilson}. {This adds further uncertainty to the absolute height of the flare ribbon with respect to the solar surface. This could potentially be constrained by inverting the region with a code that allows for the recovery of a geometrical scale, such as FIRTEZ-dz \citep{Adur2016}.}

 Similar emission heights have been produced via 1D simulations for hydrogen emission in down-flowing chromospheric compression, for example, the 5F11 RADYN flare reported in \citet{2022KowalskiBroad} and the HYDRO2GEN simulations of \citet{2018Druett}. The recent multi-dimensional simulations of \citet{2024DruettAMRVAC} also show plasma with characteristic temperatures for emission in these lines bunched around heights of 500-1000~km in their most powerful flare simulations.

\subsection{Outlook}
Flare ribbons are highly complicated and dynamic phenomena with many rapidly evolving sub-structures that require observations at high spatial and temporal resolution to be captured. The strong variation in the presented spectral features of these sub-structures shows that care should be taken when lower-resolution or average spectra are produced. 

This study provides prospective and illustrative (albeit not definitive) interpretations of the existing range among flare ribbon profiles. This work will be followed up by statistical, inversion-based, and simulated investigations to bolster the evidence provided here from the imaging and qualitative spectral analysis. 

\begin{acknowledgements}
We thank Ioannis Kontogiannis and Carsten Denker for their feedback on the manuscript text, and the anonymous referee for their valuable suggestions during the peer-review process. 
M.D. is supported by FWO project G0B4521N and also received funding from the European Research Council (ERC) under the European Union Horizon 2020 research and innovation program (grant agreement No. 833251 PROMINENT ERC-ADG 2018).
The Swedish 1-meter Solar Telescope is operated on the island of La Palma by the Institute for Solar Physics of Stockholm University in the Spanish Observatorio del Roque de los Muchachos of the Instituto de Astrof\'isica de Canarias. The Institute for Solar Physics was supported by a grant for research infrastructures of national importance from the Swedish Research Council (registration number 2017-00625).
This research has made use of NASA's Astrophysics Data System (ADS) bibliographic services. 
We acknowledge the community efforts devoted to the development of the following open-source packages that were used in this work: numpy (\href{http:\\numpy.org}{numpy.org}), matplotlib (\href{http:\\matplotlib.org}{matplotlib.org}), and astropy (\href{http:\\astropy.org}{astropy.org}).
We extensively used the CRISPEX analysis tool \citep{Gregal12}, the ISPy  \citep{ISPy2021} and CRISPy \citep{pietrow19} libraries, and SOAImage DS9 \citep{2003DS9} for data visualization. 

\end{acknowledgements}

\bibliographystyle{aa}
\bibliography{ref}

\begin{thebibliography}{111}
\expandafter\ifx\csname natexlab\endcsname\relax\def\natexlab#1{#1}\fi

\bibitem[{{Allred} {et~al.}(2005){Allred}, {Hawley}, {Abbett}, \&
  {Carlsson}}]{2005Allred}
{Allred}, J.~C., {Hawley}, S.~L., {Abbett}, W.~P., \& {Carlsson}, M. 2005,
  \href{http://dx.doi.org/10.1086/431751}{\color{magenta}\apj},
  \href{https://ui.adsabs.harvard.edu/abs/2005ApJ...630..573A}{630, 573}

\bibitem[{{Allred} \& {Kowalski}(2016)}]{2016KowalskiStark}
{Allred}, J.~C. \& {Kowalski}, A.~F. 2016, in AGU Fall Meeting Abstracts,
  \href{https://ui.adsabs.harvard.edu/abs/2016AGUFMSH43E..04A}{SH43E--04}

\bibitem[{{Allred} {et~al.}(2015){Allred}, {Kowalski}, \&
  {Carlsson}}]{2015Allred}
{Allred}, J.~C., {Kowalski}, A.~F., \& {Carlsson}, M. 2015,
  \href{http://dx.doi.org/10.1088/0004-637X/809/1/104}{\color{magenta}\apj},
  \href{https://ui.adsabs.harvard.edu/abs/2015ApJ...809..104A}{809, 104}

\bibitem[{{Asai} {et~al.}(2012){Asai}, {Ichimoto}, {Kita}, {Kurokawa}, \&
  {Shibata}}]{2012Asai}
{Asai}, A., {Ichimoto}, K., {Kita}, R., {Kurokawa}, H., \& {Shibata}, K. 2012,
  \href{http://dx.doi.org/10.1093/pasj/64.1.20}{\color{magenta}\pasj},
  \href{https://ui.adsabs.harvard.edu/abs/2012PASJ...64...20A}{64, 20}

\bibitem[{{Baker}(1970)}]{Baker1970}
{Baker}, D.~M. 1970, in Observation and Prediction of Solar Activity, Am. Inst.
  Aeronaut. Astronaut. Conf.,
  \href{https://ui.adsabs.harvard.edu/abs/1970aiaa.conf.1370B}{1370}

\bibitem[{{Berlicki} {et~al.}(2008){Berlicki}, {Heinzel}, {Schmieder}, \&
  {Li}}]{Berlicki2008}
{Berlicki}, A., {Heinzel}, P., {Schmieder}, B., \& {Li}, H. 2008,
  \href{http://dx.doi.org/10.1051/0004-6361:200809957}{\color{magenta}\aap},
  \href{https://ui.adsabs.harvard.edu/abs/2008A&A...490..315B}{490, 315}

\bibitem[{{Berlicki} {et~al.}(2005){Berlicki}, {Heinzel}, {Schmieder}, {Mein},
  \& {Mein}}]{Berlicki05}
{Berlicki}, A., {Heinzel}, P., {Schmieder}, B., {Mein}, P., \& {Mein}, N. 2005,
  \href{http://dx.doi.org/10.1051/0004-6361:20041293}{\color{magenta}\aap},
  \href{https://ui.adsabs.harvard.edu/abs/2005A&A...430..679B}{430, 679}

\bibitem[{{Brown}(1971)}]{1971Brown}
{Brown}, J.~C. 1971,
  \href{http://dx.doi.org/10.1007/BF00149070}{\color{magenta}\solphys},
  \href{https://ui.adsabs.harvard.edu/abs/1971SoPh...18..489B}{18, 489}

\bibitem[{de~la Cruz~Rodr{\'{\i}}guez {et~al.}(2019)de~la
  Cruz~Rodr{\'{\i}}guez, Leenaarts, Danilovic, \& Uitenbroek}]{Jaime19}
de~la Cruz~Rodr{\'{\i}}guez, J., Leenaarts, J., Danilovic, S., \& Uitenbroek,
  H. 2019,
  \href{http://dx.doi.org/10.1051/0004-6361/201834464}{\color{magenta}A\&A},
  623, 623

\bibitem[{{de la Cruz Rodr{\'\i}guez} {et~al.}(2015){de la Cruz
  Rodr{\'\i}guez}, {L{\"o}fdahl}, {S{\"u}tterlin}, {Hillberg}, \& {Rouppe van
  der Voort}}]{jaime15}
{de la Cruz Rodr{\'\i}guez}, J., {L{\"o}fdahl}, M.~G., {S{\"u}tterlin}, P.,
  {Hillberg}, T., \& {Rouppe van der Voort}, L. 2015,
  \href{http://dx.doi.org/10.1051/0004-6361/201424319}{\color{magenta}\aap},
  \href{https://ui.adsabs.harvard.edu/abs/2015A&A...573A..40D}{573, A40}

\bibitem[{{De Pontieu} {et~al.}(2021){De Pontieu}, {Polito}, {Hansteen},
  {Testa}, {Reeves}, {Antolin}, {N{\'o}brega-Siverio}, {Kowalski},
  {Martinez-Sykora}, {Carlsson}, {McIntosh}, {Liu}, {Daw}, \&
  {Kankelborg}}]{Bart21}
{De Pontieu}, B., {Polito}, V., {Hansteen}, V., {et~al.} 2021,
  \href{http://dx.doi.org/10.1007/s11207-021-01826-0}{\color{magenta}\solphys},
  \href{https://ui.adsabs.harvard.edu/abs/2021SoPh..296...84D}{296, 84}

\bibitem[{{De Pontieu} {et~al.}(2014){De Pontieu}, {Title}, {Lemen}, {Kushner},
  {Akin}, {Allard}, {Berger}, {Boerner}, {Cheung}, {Chou}, {Drake}, {Duncan},
  {Freeland}, {Heyman}, {Hoffman}, {Hurlburt}, {Lindgren}, {Mathur}, {Rehse},
  {Sabolish}, {Seguin}, {Schrijver}, {Tarbell}, {W{\"u}lser}, {Wolfson},
  {Yanari}, {Mudge}, {Nguyen-Phuc}, {Timmons}, {van Bezooijen}, {Weingrod},
  {Brookner}, {Butcher}, {Dougherty}, {Eder}, {Knagenhjelm}, {Larsen},
  {Mansir}, {Phan}, {Boyle}, {Cheimets}, {DeLuca}, {Golub}, {Gates}, {Hertz},
  {McKillop}, {Park}, {Perry}, {Podgorski}, {Reeves}, {Saar}, {Testa}, {Tian},
  {Weber}, {Dunn}, {Eccles}, {Jaeggli}, {Kankelborg}, {Mashburn}, {Pust},
  {Springer}, {Carvalho}, {Kleint}, {Marmie}, {Mazmanian}, {Pereira}, {Sawyer},
  {Strong}, {Worden}, {Carlsson}, {Hansteen}, {Leenaarts}, {Wiesmann},
  {Aloise}, {Chu}, {Bush}, {Scherrer}, {Brekke}, {Martinez-Sykora}, {Lites},
  {McIntosh}, {Uitenbroek}, {Okamoto}, {Gummin}, {Auker}, {Jerram}, {Pool}, \&
  {Waltham}}]{iris_2014}
{De Pontieu}, B., {Title}, A.~M., {Lemen}, J.~R., {et~al.} 2014,
  \href{http://dx.doi.org/10.1007/s11207-014-0485-y}{\color{magenta}\solphys},
  \href{https://ui.adsabs.harvard.edu/abs/2014SoPh..289.2733D}{289, 2733}

\bibitem[{{Deng} {et~al.}(2013){Deng}, {Tritschler}, {Jing}, {Chen}, {Liu},
  {Reardon}, {Denker}, {Xu}, \& {Wang}}]{2013DengIBISFlare}
{Deng}, N., {Tritschler}, A., {Jing}, J., {et~al.} 2013,
  \href{http://dx.doi.org/10.1088/0004-637X/769/2/112}{\color{magenta}\apj},
  \href{https://ui.adsabs.harvard.edu/abs/2013ApJ...769..112D}{769, 112}

\bibitem[{{D\'iaz Baso} {et~al.}(2021){D\'iaz Baso}, {Vissers}, {Calvo},
  {Pietrow}, {Yadav}, {de la Cruz Rodr{\'\i}guez}, \& {Zivadinovic}}]{ISPy2021}
{D\'iaz Baso}, C., {Vissers}, G., {Calvo}, F., {et~al.} 2021, in Zenodo
  software package, Vol.~56,
  \href{https://ui.adsabs.harvard.edu/abs/2021zndo...5608441D}{5608441}

\bibitem[{{Dodson} {et~al.}(1956){Dodson}, {Hedeman}, \&
  {McMath}}]{1956Dodson_flares}
{Dodson}, H.~W., {Hedeman}, E.~R., \& {McMath}, R.~R. 1956,
  \href{http://dx.doi.org/10.1086/190027}{\color{magenta}\apjs},
  \href{https://ui.adsabs.harvard.edu/abs/1956ApJS....2..241D}{2, 241}

\bibitem[{{Druett} {et~al.}(2023{\natexlab{a}}){Druett}, {Ruan}, \&
  {Keppens}}]{2024DruettAMRVAC}
{Druett}, M., {Ruan}, W., \& {Keppens}, R. 2023{\natexlab{a}},
  \href{https://ui.adsabs.harvard.edu/abs/2023arXiv231009939D}{\href{http://dx.doi.org/10.48550/arXiv.2310.09939}{\color{magenta}arXiv
  e-prints}, arXiv:2310.09939}

\bibitem[{{Druett} {et~al.}(2017){Druett}, {Scullion}, {Zharkova}, {Matthews},
  {Zharkov}, \& {Rouppe van der Voort}}]{2017Druett}
{Druett}, M., {Scullion}, E., {Zharkova}, V., {et~al.} 2017,
  \href{http://dx.doi.org/10.1038/ncomms15905}{\color{magenta}Nature
  Communications},
  \href{https://ui.adsabs.harvard.edu/abs/2017NatCo...815905D}{8, 15905}

\bibitem[{{Druett} {et~al.}(2024){Druett}, {Kontogiannis}, {Pietrow}, {Dineva},
  {Keppens}, \& {Denker}}]{2024DruettIRIS}
{Druett}, M.~K., {Kontogiannis}, I., {Pietrow}, A.~G.~M., {et~al.} 2024, arXiv
  e-prints

\bibitem[{{Druett} {et~al.}(2021){Druett}, {Pietrow}, \&
  {Vissers}}]{DruettPoster}
{Druett}, M.~K., {Pietrow}, A.~G.~M., \& {Vissers}, G.~J.~M. 2021, in SolFER
  Spring 2021 Meeting,
  \href{https://ui.adsabs.harvard.edu/abs/2021solf.confE...1D}{E1}

\bibitem[{{Druett} {et~al.}(2022){Druett}, {Pietrow}, {Vissers}, {Robustini},
  \& {Calvo}}]{2022druettcoco}
{Druett}, M.~K., {Pietrow}, A. G.~M., {Vissers}, G. J.~M., {Robustini}, C., \&
  {Calvo}, F. 2022,
  \href{http://dx.doi.org/10.1093/rasti/rzac003}{\color{magenta}RASTI},
  \href{https://ui.adsabs.harvard.edu/abs/2022RASTI...1...29D}{1, 29}

\bibitem[{{Druett} {et~al.}(2023{\natexlab{b}}){Druett}, {Ruan}, \&
  {Keppens}}]{2023DruettAMRVAC}
{Druett}, M.~K., {Ruan}, W., \& {Keppens}, R. 2023{\natexlab{b}},
  \href{https://ui.adsabs.harvard.edu/abs/2023arXiv231011226D}{arXiv e-prints,
  arXiv:2310.11226}

\bibitem[{{Druett} \& {Zharkova}(2018)}]{2018Druett}
{Druett}, M.~K. \& {Zharkova}, V.~V. 2018,
  \href{http://dx.doi.org/10.1051/0004-6361/201731053}{\color{magenta}\aap},
  \href{https://ui.adsabs.harvard.edu/abs/2018A&A...610A..68D}{610, A68}

\bibitem[{{Druett} \& {Zharkova}(2019)}]{2019Druett}
{Druett}, M.~K. \& {Zharkova}, V.~V. 2019,
  \href{http://dx.doi.org/10.1051/0004-6361/201732427}{\color{magenta}\aap},
  \href{https://ui.adsabs.harvard.edu/abs/2019A&A...623A..20D}{623, A20}

\bibitem[{{Emslie}(1978)}]{1978Emslie}
{Emslie}, A.~G. 1978,
  \href{http://dx.doi.org/10.1086/156371}{\color{magenta}\apj},
  \href{https://ui.adsabs.harvard.edu/abs/1978ApJ...224..241E}{224, 241}

\bibitem[{{Fletcher} {et~al.}(2011){Fletcher}, {Dennis}, {Hudson}, {Krucker},
  {Phillips}, {Veronig}, {Battaglia}, {Bone}, {Caspi}, {Chen}, {Gallagher},
  {Grigis}, {Ji}, {Liu}, {Milligan}, \& {Temmer}}]{2011Fletcher}
{Fletcher}, L., {Dennis}, B.~R., {Hudson}, H.~S., {et~al.} 2011,
  \href{http://dx.doi.org/10.1007/s11214-010-9701-8}{\color{magenta}\ssr},
  \href{https://ui.adsabs.harvard.edu/abs/2011SSRv..159...19F}{159, 19}

\bibitem[{{Harvey}(1971)}]{1971Harvey_flare_kernels}
{Harvey}, K.~L. 1971,
  \href{http://dx.doi.org/10.1007/BF00162485}{\color{magenta}\solphys},
  \href{https://ui.adsabs.harvard.edu/abs/1971SoPh...16..423H}{16, 423}

\bibitem[{{Huang} {et~al.}(2019){Huang}, {Xu}, {Sadykov}, {Jing}, \&
  {Wang}}]{2019HuangFlareMghkRADYN}
{Huang}, N., {Xu}, Y., {Sadykov}, V.~M., {Jing}, J., \& {Wang}, H. 2019,
  \href{http://dx.doi.org/10.3847/2041-8213/ab2330}{\color{magenta}\apjl},
  \href{https://ui.adsabs.harvard.edu/abs/2019ApJ...878L..15H}{878, L15}

\bibitem[{{Ichimoto} \& {Kurokawa}(1984)}]{Ichomoto1984}
{Ichimoto}, K. \& {Kurokawa}, H. 1984,
  \href{http://dx.doi.org/10.1007/BF00156656}{\color{magenta}\solphys},
  \href{https://ui.adsabs.harvard.edu/abs/1984SoPh...93..105I}{93, 105}

\bibitem[{{Joye} \& {Mandel}(2003)}]{2003DS9}
{Joye}, W.~A. \& {Mandel}, E. 2003, in ASP Conf. Ser., Vol. 295, Astronomical
  Data Analysis Software and Systems XII, ed. H.~E. {Payne}, R.~I.
  {Jedrzejewski}, \& R.~N. {Hook},
  \href{http://adsabs.harvard.edu/abs/2003ASPC..295..489J}{489}

\bibitem[{{Keppens} {et~al.}(2023){Keppens}, {Popescu Braileanu}, {Zhou},
  {Ruan}, {Xia}, {Guo}, {Claes}, \& {Bacchini}}]{2023KeppensAMRVAC3}
{Keppens}, R., {Popescu Braileanu}, B., {Zhou}, Y., {et~al.} 2023,
  \href{http://dx.doi.org/10.1051/0004-6361/202245359}{\color{magenta}\aap},
  \href{https://ui.adsabs.harvard.edu/abs/2023A&A...673A..66K}{673, A66}

\bibitem[{{Kerr}(2022)}]{Kerr2022}
{Kerr}, G.~S. 2022,
  \href{http://dx.doi.org/10.3389/fspas.2022.1060856}{\color{magenta}Frontiers
  in Astronomy and Space Sciences},
  \href{https://ui.adsabs.harvard.edu/abs/2022FrASS...960856K}{9, 1060856}

\bibitem[{{Kerr} {et~al.}(2019{\natexlab{a}}){Kerr}, {Allred}, \&
  {Carlsson}}]{2019KerrFlareMgIIpartI}
{Kerr}, G.~S., {Allred}, J.~C., \& {Carlsson}, M. 2019{\natexlab{a}},
  \href{http://dx.doi.org/10.3847/1538-4357/ab3c24}{\color{magenta}\apj},
  \href{https://ui.adsabs.harvard.edu/abs/2019ApJ...883...57K}{883, 57}

\bibitem[{{Kerr} {et~al.}(2023{\natexlab{a}}){Kerr}, {Allred}, {Kowalski},
  {Milligan}, {Hudson}, {Zambrana Prado}, {Kucera}, \&
  {Brosius}}]{2023KerrOrrallZirker}
{Kerr}, G.~S., {Allred}, J.~C., {Kowalski}, A.~F., {et~al.} 2023{\natexlab{a}},
  \href{http://dx.doi.org/10.3847/1538-4357/acb92a}{\color{magenta}\apj},
  \href{https://ui.adsabs.harvard.edu/abs/2023ApJ...945..118K}{945, 118}

\bibitem[{{Kerr} {et~al.}(2020){Kerr}, {Allred}, \& {Polito}}]{2020Kerr}
{Kerr}, G.~S., {Allred}, J.~C., \& {Polito}, V. 2020,
  \href{http://dx.doi.org/10.3847/1538-4357/abaa46}{\color{magenta}\apj},
  \href{https://ui.adsabs.harvard.edu/abs/2020ApJ...900...18K}{900, 18}

\bibitem[{{Kerr} {et~al.}(2019{\natexlab{b}}){Kerr}, {Carlsson}, \&
  {Allred}}]{2019KerrFlareMgIIpartII}
{Kerr}, G.~S., {Carlsson}, M., \& {Allred}, J.~C. 2019{\natexlab{b}},
  \href{http://dx.doi.org/10.3847/1538-4357/ab48ea}{\color{magenta}\apj},
  \href{https://ui.adsabs.harvard.edu/abs/2019ApJ...885..119K}{885, 119}

\bibitem[{{Kerr} {et~al.}(2019{\natexlab{c}}){Kerr}, {Carlsson}, {Allred},
  {Young}, \& {Daw}}]{2019KerrFlareSiIV}
{Kerr}, G.~S., {Carlsson}, M., {Allred}, J.~C., {Young}, P.~R., \& {Daw}, A.~N.
  2019{\natexlab{c}},
  \href{http://dx.doi.org/10.3847/1538-4357/aaf46e}{\color{magenta}\apj},
  \href{https://ui.adsabs.harvard.edu/abs/2019ApJ...871...23K}{871, 23}

\bibitem[{{Kerr} {et~al.}(2023{\natexlab{b}}){Kerr}, {Kowalski}, {Allred},
  {Daw}, \& {Kane}}]{2023KerrWidths}
{Kerr}, G.~S., {Kowalski}, A.~F., {Allred}, J.~C., {Daw}, A.~N., \& {Kane},
  M.~R. 2023{\natexlab{b}},
  \href{https://ui.adsabs.harvard.edu/abs/2023arXiv231007111K}{\href{http://dx.doi.org/10.48550/arXiv.2310.07111}{\color{magenta}arXiv
  e-prints}, arXiv:2310.07111}

\bibitem[{{Kerr} {et~al.}(2021){Kerr}, {Xu}, {Allred}, {Polito}, {Sadykov},
  {Huang}, \& {Wang}}]{2021KerrFlareHeDimming}
{Kerr}, G.~S., {Xu}, Y., {Allred}, J.~C., {et~al.} 2021,
  \href{http://dx.doi.org/10.3847/1538-4357/abf42d}{\color{magenta}\apj},
  \href{https://ui.adsabs.harvard.edu/abs/2021ApJ...912..153K}{912, 153}

\bibitem[{{Kleint} {et~al.}(2020){Kleint}, {Berkefeld}, {Esteves}, {Sonner},
  {Volkmer}, {Gerber}, {Kr{\"a}mer}, {Grassin}, \& {Berdyugina}}]{Kleint2020}
{Kleint}, L., {Berkefeld}, T., {Esteves}, M., {et~al.} 2020,
  \href{http://dx.doi.org/10.1051/0004-6361/202038208}{\color{magenta}\aap},
  \href{https://ui.adsabs.harvard.edu/abs/2020A&A...641A..27K}{641, A27}

\bibitem[{{Kosugi} {et~al.}(2007){Kosugi}, {Matsuzaki}, {Sakao}, {Shimizu},
  {Sone}, {Tachikawa}, {Hashimoto}, {Minesugi}, {Ohnishi}, {Yamada}, {Tsuneta},
  {Hara}, {Ichimoto}, {Suematsu}, {Shimojo}, {Watanabe}, {Shimada}, {Davis},
  {Hill}, {Owens}, {Title}, {Culhane}, {Harra}, {Doschek}, \&
  {Golub}}]{2007KosugiHinode}
{Kosugi}, T., {Matsuzaki}, K., {Sakao}, T., {et~al.} 2007,
  \href{http://dx.doi.org/10.1007/s11207-007-9014-6}{\color{magenta}\solphys},
  \href{https://ui.adsabs.harvard.edu/abs/2007SoPh..243....3K}{243, 3}

\bibitem[{{Kowalski} {et~al.}(2022){Kowalski}, {Allred}, {Carlsson}, {Kerr},
  {Tremblay}, {Namekata}, {Kuridze}, \& {Uitenbroek}}]{2022KowalskiBroad}
{Kowalski}, A.~F., {Allred}, J.~C., {Carlsson}, M., {et~al.} 2022,
  \href{http://dx.doi.org/10.3847/1538-4357/ac5174}{\color{magenta}\apj},
  \href{https://ui.adsabs.harvard.edu/abs/2022ApJ...928..190K}{928, 190}

\bibitem[{{Kowalski} {et~al.}(2017{\natexlab{a}}){Kowalski}, {Allred}, {Daw},
  {Cauzzi}, \& {Carlsson}}]{2017Kowalskidown}
{Kowalski}, A.~F., {Allred}, J.~C., {Daw}, A., {Cauzzi}, G., \& {Carlsson}, M.
  2017{\natexlab{a}},
  \href{http://dx.doi.org/10.3847/1538-4357/836/1/12}{\color{magenta}\apj},
  \href{https://ui.adsabs.harvard.edu/abs/2017ApJ...836...12K}{836, 12}

\bibitem[{{Kowalski} {et~al.}(2017{\natexlab{b}}){Kowalski}, {Allred},
  {Uitenbroek}, {Tremblay}, {Brown}, {Carlsson}, {Osten}, {Wisniewski}, \&
  {Hawley}}]{2017KowalskiBroad}
{Kowalski}, A.~F., {Allred}, J.~C., {Uitenbroek}, H., {et~al.}
  2017{\natexlab{b}},
  \href{http://dx.doi.org/10.3847/1538-4357/aa603e}{\color{magenta}\apj},
  \href{https://ui.adsabs.harvard.edu/abs/2017ApJ...837..125K}{837, 125}

\bibitem[{{Kowalski} {et~al.}(2019){Kowalski}, {Butler}, {Daw}, {Fletcher},
  {Allred}, {De Pontieu}, {Kerr}, \& {Cauzzi}}]{2019Kowalski}
{Kowalski}, A.~F., {Butler}, E., {Daw}, A.~N., {et~al.} 2019,
  \href{http://dx.doi.org/10.3847/1538-4357/ab1f8b}{\color{magenta}\apj},
  \href{https://ui.adsabs.harvard.edu/abs/2019ApJ...878..135K}{878, 135}

\bibitem[{{Krucker} {et~al.}(2015){Krucker}, {Saint-Hilaire}, {Hudson},
  {Haberreiter}, {Martinez-Oliveros}, {Fivian}, {Hurford}, {Kleint},
  {Battaglia}, {Kuhar}, \& {Arnold}}]{2015Krucker}
{Krucker}, S., {Saint-Hilaire}, P., {Hudson}, H.~S., {et~al.} 2015,
  \href{http://dx.doi.org/10.1088/0004-637X/802/1/19}{\color{magenta}\apj},
  \href{https://ui.adsabs.harvard.edu/abs/2015ApJ...802...19K}{802, 19}

\bibitem[{{Kuridze} {et~al.}(2020){Kuridze}, {Mathioudakis}, {Heinzel}, {Koza},
  {Morgan}, {Oliver}, {Kowalski}, \& {Allred}}]{2020KuridzeLimbFlare}
{Kuridze}, D., {Mathioudakis}, M., {Heinzel}, P., {et~al.} 2020,
  \href{http://dx.doi.org/10.3847/1538-4357/ab9603}{\color{magenta}\apj},
  \href{https://ui.adsabs.harvard.edu/abs/2020ApJ...896..120K}{896, 120}

\bibitem[{{Leenaarts} {et~al.}(2007){Leenaarts}, {Carlsson}, {Hansteen}, \&
  {Rutten}}]{2007Leenaarts}
{Leenaarts}, J., {Carlsson}, M., {Hansteen}, V., \& {Rutten}, R.~J. 2007,
  \href{http://dx.doi.org/10.1051/0004-6361:20078161}{\color{magenta}\aap},
  \href{https://ui.adsabs.harvard.edu/abs/2007A&A...473..625L}{473, 625}

\bibitem[{{Leenaarts} {et~al.}(2012){Leenaarts}, {Pereira}, \&
  {Uitenbroek}}]{jorrit12}
{Leenaarts}, J., {Pereira}, T., \& {Uitenbroek}, H. 2012,
  \href{http://dx.doi.org/10.1051/0004-6361/201219394}{\color{magenta}ASP Conf.
  Ser.}, \href{https://ui.adsabs.harvard.edu/abs/2012A&A...543A.109L}{543,
  A109}

\bibitem[{{Lemen} {et~al.}(2012){Lemen}, {Title}, {Akin}, {Boerner}, {Chou},
  {Drake}, {Duncan}, {Edwards}, {Friedlaender}, {Heyman}, {Hurlburt}, {Katz},
  {Kushner}, {Levay}, {Lindgren}, {Mathur}, {McFeaters}, {Mitchell}, {Rehse},
  {Schrijver}, {Springer}, {Stern}, {Tarbell}, {Wuelser}, {Wolfson}, {Yanari},
  {Bookbinder}, {Cheimets}, {Caldwell}, {Deluca}, {Gates}, {Golub}, {Park},
  {Podgorski}, {Bush}, {Scherrer}, {Gummin}, {Smith}, {Auker}, {Jerram},
  {Pool}, {Soufli}, {Windt}, {Beardsley}, {Clapp}, {Lang}, \&
  {Waltham}}]{Lemen2012AIASDO}
{Lemen}, J.~R., {Title}, A.~M., {Akin}, D.~J., {et~al.} 2012,
  \href{http://dx.doi.org/10.1007/s11207-011-9776-8}{\color{magenta}\solphys},
  \href{https://ui.adsabs.harvard.edu/abs/2012SoPh..275...17L}{275, 17}

\bibitem[{{L{\"o}fdahl}(2002)}]{mats02}
{L{\"o}fdahl}, M.~G. 2002, in Proc SPIE, Vol. 4792, Image Reconstruction from
  Incomplete Data II, ed. P.~J. {Bones}, M.~A. {Fiddy}, \& R.~P. {Millane},
  \href{https://ui.adsabs.harvard.edu/abs/2002SPIE.4792..146L}{146--155}

\bibitem[{{L{\"o}fdahl} {et~al.}(2021){L{\"o}fdahl}, {Hillberg}, {de la Cruz
  Rodr{\'i}guez}, {Vissers}, {Andriienko}, {Scharmer}, {Haugan}, \&
  {Fredvik}}]{mats21}
{L{\"o}fdahl}, M.~G., {Hillberg}, T., {de la Cruz Rodr{\'i}guez}, J., {et~al.}
  2021,
  \href{http://dx.doi.org/10.1051/0004-6361/202141326}{\color{magenta}A\&A},
  653, 653

\bibitem[{{L{\"o}ptien} {et~al.}(2018){L{\"o}ptien}, {Lagg}, {van Noort}, \&
  {Solanki}}]{Loptien2018}
{L{\"o}ptien}, B., {Lagg}, A., {van Noort}, M., \& {Solanki}, S.~K. 2018,
  \href{http://dx.doi.org/10.1051/0004-6361/201833571}{\color{magenta}\aap},
  \href{https://ui.adsabs.harvard.edu/abs/2018A&A...619A..42L}{619, A42}

\bibitem[{Machol {et~al.}(2020)Machol, Eparvier, Viereck, Woodraska, Snow,
  Thiemann, Woods, McClintock, Mueller, Eden, Meisner, Codrescu, Bouwer, \&
  Reinard}]{Machol2020}
Machol, J.~L., Eparvier, F.~G., Viereck, R.~A., {et~al.} 2020, in The {GOES}-R
  Series (Elsevier), 233--242

\bibitem[{{Maehara} {et~al.}(2015){Maehara}, {Shibayama}, {Notsu}, {Notsu},
  {Honda}, {Nogami}, \& {Shibata}}]{Maehara15}
{Maehara}, H., {Shibayama}, T., {Notsu}, Y., {et~al.} 2015,
  \href{http://dx.doi.org/10.1186/s40623-015-0217-z}{\color{magenta}Earth,
  Planets and Space},
  \href{https://ui.adsabs.harvard.edu/abs/2015EP&S...67...59M}{67, 59}

\bibitem[{{Mart{\'\i}nez Oliveros} {et~al.}(2012){Mart{\'\i}nez Oliveros},
  {Hudson}, {Hurford}, {Krucker}, {Lin}, {Lindsey}, {Couvidat}, {Schou}, \&
  {Thompson}}]{2012MartinezOliverosHXRFlare}
{Mart{\'\i}nez Oliveros}, J.-C., {Hudson}, H.~S., {Hurford}, G.~J., {et~al.}
  2012,
  \href{http://dx.doi.org/10.1088/2041-8205/753/2/L26}{\color{magenta}\apjl},
  \href{https://ui.adsabs.harvard.edu/abs/2012ApJ...753L..26M}{753, L26}

\bibitem[{{Molnar} {et~al.}(2019){Molnar}, {Reardon}, {Chai}, {Gary},
  {Uitenbroek}, {Cauzzi}, \& {Cranmer}}]{molnar2019}
{Molnar}, M.~E., {Reardon}, K.~P., {Chai}, Y., {et~al.} 2019,
  \href{http://dx.doi.org/10.3847/1538-4357/ab2ba3}{\color{magenta}\apj},
  \href{https://ui.adsabs.harvard.edu/abs/2019ApJ...881...99M}{881, 99}

\bibitem[{{Monson} {et~al.}(2021){Monson}, {Mathioudakis}, {Reid}, {Milligan},
  \& {Kuridze}}]{Monson2021}
{Monson}, A.~J., {Mathioudakis}, M., {Reid}, A., {Milligan}, R., \& {Kuridze},
  D. 2021,
  \href{http://dx.doi.org/10.3847/1538-4357/abfda8}{\color{magenta}\apj},
  \href{https://ui.adsabs.harvard.edu/abs/2021ApJ...915...16M}{915, 16}

\bibitem[{{Morosin} {et~al.}(2020){Morosin}, {de la Cruz Rodriguez}, {Vissers},
  \& {Yadav}}]{roberta2020}
{Morosin}, R., {de la Cruz Rodriguez}, J., {Vissers}, G.~J.~M., \& {Yadav}, R.
  2020, \href{https://ui.adsabs.harvard.edu/abs/2020arXiv200614487M}{arXiv
  e-prints, arXiv:2006.14487}

\bibitem[{{Orrall} \& {Zirker}(1976)}]{1976OrrallZirker}
{Orrall}, F.~Q. \& {Zirker}, J.~B. 1976,
  \href{http://dx.doi.org/10.1086/154642}{\color{magenta}\apj},
  \href{https://ui.adsabs.harvard.edu/abs/1976ApJ...208..618O}{208, 618}

\bibitem[{{Osborne} \& {Fletcher}(2022)}]{2022OsborneFlareKernels}
{Osborne}, C. M.~J. \& {Fletcher}, L. 2022,
  \href{http://dx.doi.org/10.1093/mnras/stac2570}{\color{magenta}\mnras},
  \href{https://ui.adsabs.harvard.edu/abs/2022MNRAS.516.6066O}{516, 6066}

\bibitem[{{Otsu} \& {Asai}(2024)}]{Otsu24}
{Otsu}, T. \& {Asai}, A. 2024,
  \href{https://ui.adsabs.harvard.edu/abs/2024arXiv240200589O}{\href{http://dx.doi.org/10.48550/arXiv.2402.00589}{\color{magenta}arXiv
  e-prints}, arXiv:2402.00589}

\bibitem[{{Otsu} {et~al.}(2022){Otsu}, {Asai}, {Ichimoto}, {Ishii}, \&
  {Namekata}}]{2022Otsu}
{Otsu}, T., {Asai}, A., {Ichimoto}, K., {Ishii}, T.~T., \& {Namekata}, K. 2022,
  \href{http://dx.doi.org/10.3847/1538-4357/ac9730}{\color{magenta}\apj},
  \href{https://ui.adsabs.harvard.edu/abs/2022ApJ...939...98O}{939, 98}

\bibitem[{{Parker}(1963)}]{1963Parker}
{Parker}, E.~N. 1963,
  \href{http://dx.doi.org/10.1086/190087}{\color{magenta}\apjs},
  \href{https://ui.adsabs.harvard.edu/abs/1963ApJS....8..177P}{8, 177}

\bibitem[{{Pastor Yabar} {et~al.}(2019){Pastor Yabar}, {Borrero}, \& {Ruiz
  Cobo}}]{Adur2016}
{Pastor Yabar}, A., {Borrero}, J.~M., \& {Ruiz Cobo}, B. 2019,
  \href{http://dx.doi.org/10.1051/0004-6361/201935692}{\color{magenta}\aap},
  \href{https://ui.adsabs.harvard.edu/abs/2019A&A...629A..24P}{629, A24}

\bibitem[{{Pesnell} {et~al.}(2012){Pesnell}, {Thompson}, \&
  {Chamberlin}}]{2012PesnellSDO}
{Pesnell}, W.~D., {Thompson}, B.~J., \& {Chamberlin}, P.~C. 2012,
  \href{http://dx.doi.org/10.1007/s11207-011-9841-3}{\color{magenta}\solphys},
  \href{https://ui.adsabs.harvard.edu/abs/2012SoPh..275....3P}{275, 3}

\bibitem[{{Petschek}(1964)}]{1964Petschek}
{Petschek}, H.~E. 1964, in NASA Special Publication, Vol.~50, 425

\bibitem[{{Pietras} {et~al.}(2022){Pietras}, {Falewicz}, {Siarkowski}, {Bicz},
  \& {Pre{\'s}}}]{Pietras2022}
{Pietras}, M., {Falewicz}, R., {Siarkowski}, M., {Bicz}, K., \& {Pre{\'s}}, P.
  2022, \href{http://dx.doi.org/10.3847/1538-4357/ac8352}{\color{magenta}\apj},
  \href{https://ui.adsabs.harvard.edu/abs/2022ApJ...935..143P}{935, 143}

\bibitem[{{Pietrow}(2019)}]{pietrow19}
{Pietrow}, A.~G.~M. 2019, {AlexPietrow/CRISpy 0.31}

\bibitem[{{Pietrow}(2022)}]{Pietrowthesis}
{Pietrow}, A.~G.~M. 2022,
  \href{https://ui.adsabs.harvard.edu/abs/2022PhDT.........3P}{{Physical
  properties of chromospheric features: Plage, peacock jets, and calibrating it
  all}}, PhD thesis, Stockholm University

\bibitem[{{Pietrow} {et~al.}(2024){Pietrow}, {Cretignier}, {Druett},
  {Alvarado-G{\'o}mez}, {Hofmeister}, {Verma}, {Kamlah}, {Baratella},
  {Amazo-G{\'o}mez}, {Kontogiannis}, {Dineva}, {Warmuth}, {Denker},
  {Poppenhaeger}, {Andriienko}, {Dumusque}, \& {L{\"o}fdahl}}]{Pietrow24harps}
{Pietrow}, A.~G.~M., {Cretignier}, M., {Druett}, M.~K., {et~al.} 2024,
  \href{http://dx.doi.org/10.1051/0004-6361/202347895}{\color{magenta}\aap},
  \href{https://ui.adsabs.harvard.edu/abs/2024A&A...682A..46P}{682, A46}

\bibitem[{{Pietrow} {et~al.}(2022){Pietrow}, {Druett}, {de la Cruz Rodriguez},
  {Calvo}, \& {Kiselman}}]{Pietrow22a}
{Pietrow}, A.~G.~M., {Druett}, M.~K., {de la Cruz Rodriguez}, J., {Calvo}, F.,
  \& {Kiselman}, D. 2022,
  \href{http://dx.doi.org/10.1051/0004-6361/202142346}{\color{magenta}\aap},
  \href{https://ui.adsabs.harvard.edu/abs/2022A&A...659A..58P}{659, A58}

\bibitem[{{Polito} {et~al.}(2023){Polito}, {Kerr}, {Xu}, {Sadykov}, \&
  {Lorincik}}]{2023PolitoRibbonLeadingEdge}
{Polito}, V., {Kerr}, G.~S., {Xu}, Y., {Sadykov}, V.~M., \& {Lorincik}, J.
  2023, \href{http://dx.doi.org/10.3847/1538-4357/acaf7c}{\color{magenta}\apj},
  \href{https://ui.adsabs.harvard.edu/abs/2023ApJ...944..104P}{944, 104}

\bibitem[{{Quinn} {et~al.}(2019){Quinn}, {Reid}, {Mathioudakis}, {Nelson},
  {Krishna Prasad}, \& {Zharkov}}]{Quinn19}
{Quinn}, S., {Reid}, A., {Mathioudakis}, M., {et~al.} 2019,
  \href{http://dx.doi.org/10.3847/1538-4357/ab2c9e}{\color{magenta}\apj},
  \href{https://ui.adsabs.harvard.edu/abs/2019ApJ...881...82Q}{881, 82}

\bibitem[{{Ruan} {et~al.}(2020){Ruan}, {Xia}, \& {Keppens}}]{2020Ruan}
{Ruan}, W., {Xia}, C., \& {Keppens}, R. 2020,
  \href{http://dx.doi.org/10.3847/1538-4357/ab93db}{\color{magenta}\apj},
  \href{https://ui.adsabs.harvard.edu/abs/2020ApJ...896...97R}{896, 97}

\bibitem[{{Rutten}(2003)}]{2003Rutten}
{Rutten}, R.~J. 2003, {Radiative Transfer in Stellar Atmospheres}

\bibitem[{{Scharmer}(2017)}]{Scharmer17}
{Scharmer}, G. 2017, in SOLARNET IV: The Physics of the Sun from the Interior
  to the Outer Atmosphere,
  \href{https://ui.adsabs.harvard.edu/abs/2017psio.confE..85S}{85}

\bibitem[{{Scharmer} {et~al.}(2003){Scharmer}, {Bjelksjo}, {Korhonen},
  {Lindberg}, \& {Petterson}}]{Scharmer03}
{Scharmer}, G.~B., {Bjelksjo}, K., {Korhonen}, T.~K., {Lindberg}, B., \&
  {Petterson}, B. 2003, in Proc. SPIE, Vol. 4853, Innovative Telescopes and
  Instrumentation for Solar Astrophysics, ed. S.~L. {Keil} \& S.~V. {Avakyan},
  \href{https://ui.adsabs.harvard.edu/abs/2003SPIE.4853..341S}{341--350}

\bibitem[{{Scharmer} {et~al.}(2008){Scharmer}, {Narayan}, {Hillberg}, {de la
  Cruz Rodriguez}, {L{\"o}fdahl}, {Kiselman}, {S{\"u}tterlin}, {van Noort}, \&
  {Lagg}}]{Scharmer08}
{Scharmer}, G.~B., {Narayan}, G., {Hillberg}, T., {et~al.} 2008,
  \href{http://dx.doi.org/10.1086/595744}{\color{magenta}\apj},
  \href{https://ui.adsabs.harvard.edu/abs/2008ApJ...689L..69S}{689, L69}

\bibitem[{{Scherrer} {et~al.}(2012){Scherrer}, {Schou}, {Bush}, {Kosovichev},
  {Bogart}, {Hoeksema}, {Liu}, {Duvall}, {Zhao}, {Title}, {Schrijver},
  {Tarbell}, \& {Tomczyk}}]{Scherrer2012HMI}
{Scherrer}, P.~H., {Schou}, J., {Bush}, R.~I., {et~al.} 2012,
  \href{http://dx.doi.org/10.1007/s11207-011-9834-2}{\color{magenta}\solphys},
  \href{https://ui.adsabs.harvard.edu/abs/2012SoPh..275..207S}{275, 207}

\bibitem[{{Schmidt} {et~al.}(2012){Schmidt}, {von der L{\"u}he}, {Volkmer},
  {Denker}, {Solanki}, {Balthasar}, {Bello Gonz{\'a}lez}, {Berkefeld},
  {Collados}, {Fischer}, {Halbgewachs}, {Heidecke}, {Hofmann}, {Kneer}, {Lagg},
  {Nicklas}, {Popow}, {Puschmann}, {Schmidt}, {Sigwarth}, {Sobotka}, {Soltau},
  {Staude}, {Strassmeier}, \& {Waldmann}}]{Schmidt12}
{Schmidt}, W., {von der L{\"u}he}, O., {Volkmer}, R., {et~al.} 2012,
  \href{http://dx.doi.org/10.1002/asna.201211725}{\color{magenta}Astronomische
  Nachrichten},
  \href{https://ui.adsabs.harvard.edu/abs/2012AN....333..796S}{333, 796}

\bibitem[{{Schmieder} {et~al.}(1990){Schmieder}, {Malherbe}, {Simnett},
  {Forbes}, \& {Tandberg-Hanssen}}]{Schmieder1990}
{Schmieder}, B., {Malherbe}, J.~M., {Simnett}, G.~M., {Forbes}, T.~G., \&
  {Tandberg-Hanssen}, E. 1990,
  \href{http://dx.doi.org/10.1086/168879}{\color{magenta}\apj},
  \href{https://ui.adsabs.harvard.edu/abs/1990ApJ...356..720S}{356, 720}

\bibitem[{{Schmit} {et~al.}(2017){Schmit}, {Griffith}, {Gunshor}, {Daniels},
  {Goodman}, \& {Lebair}}]{2017BAMS...98..681S}
{Schmit}, T.~J., {Griffith}, P., {Gunshor}, M.~M., {et~al.} 2017,
  \href{http://dx.doi.org/10.1175/BAMS-D-15-00230.1}{\color{magenta}Bull. Am.
  Meteorol. Soc.},
  \href{https://ui.adsabs.harvard.edu/abs/2017BAMS...98..681S}{98, 681}

\bibitem[{{Schmit} {et~al.}(2005){Schmit}, {Gunshor}, {Menzel}, {Gurka}, {Li},
  \& {Bachmeier}}]{2005BAMS...86.1079S}
{Schmit}, T.~J., {Gunshor}, M.~M., {Menzel}, W.~P., {et~al.} 2005,
  \href{http://dx.doi.org/10.1175/BAMS-86-8-1079}{\color{magenta}Bull. Am.
  Meteorol. Soc.},
  \href{https://ui.adsabs.harvard.edu/abs/2005BAMS...86.1079S}{86, 1079}

\bibitem[{{Schmit} {et~al.}(2019){Schmit}, {Li}, {Lee}, {Li}, {Dworak}, {Lee},
  {Bowlan}, {Gerth}, {Martin}, {Straka}, {Baggett}, \&
  {Cronce}}]{2019E&SS....6.1730S}
{Schmit}, T.~J., {Li}, J., {Lee}, S.~J., {et~al.} 2019,
  \href{http://dx.doi.org/10.1029/2019EA000729}{\color{magenta}Earth Space
  Science}, \href{https://ui.adsabs.harvard.edu/abs/2019E&SS....6.1730S}{6,
  1730}

\bibitem[{{Shibata}(1998)}]{1998Shibata}
{Shibata}, K. 1998, in Astrophysics and Space Science Library, Vol. 229,
  Observational Plasma Astrophysics : Five Years of YOHKOH and Beyond, ed.
  T.~{Watanabe} \& T.~{Kosugi},
  \href{https://ui.adsabs.harvard.edu/abs/1998ASSL..229..187S}{187}

\bibitem[{{Sim{\~o}es} {et~al.}(2024){Sim{\~o}es}, {Ara{\'u}jo}, {Valio}, \&
  {Fletcher}}]{Simoes2024}
{Sim{\~o}es}, P. J.~A., {Ara{\'u}jo}, A., {Valio}, A., \& {Fletcher}, L. 2024,
  \href{http://dx.doi.org/10.1093/mnras/stae186}{\color{magenta}\mnras}
  \href{https://ui.adsabs.harvard.edu/abs/2024MNRAS.tmp..187S}{[\eprint[arXiv]{2401.07824}]}

\bibitem[{{Sim{\~o}es} {et~al.}(2017){Sim{\~o}es}, {Kerr}, {Fletcher},
  {Hudson}, {Gim{\'e}nez de Castro}, \& {Penn}}]{2017Simoes}
{Sim{\~o}es}, P. J.~A., {Kerr}, G.~S., {Fletcher}, L., {et~al.} 2017,
  \href{http://dx.doi.org/10.1051/0004-6361/201730856}{\color{magenta}\aap},
  \href{https://ui.adsabs.harvard.edu/abs/2017A&A...605A.125S}{605, A125}

\bibitem[{{Singh} {et~al.}(2024){Singh}, {Scullion}, {Botha}, {Jeffrey},
  {Druett}, \& {et al.}}]{2023_Singh_Riblets}
{Singh}, V., {Scullion}, E., {Botha}, G., {et~al.} 2024, ApJ (Submitted)

\bibitem[{{Stokes}(1851)}]{Stokes1951}
{Stokes}, G.~G. 1851, Transactions of the Cambridge Philosophical Society,
  \href{https://ui.adsabs.harvard.edu/abs/1851TCaPS...9..399S}{9, 399}

\bibitem[{{Sturrock}(1966)}]{1966Sturrock}
{Sturrock}, P.~A. 1966,
  \href{http://dx.doi.org/10.1038/211695a0}{\color{magenta}\nat},
  \href{https://ui.adsabs.harvard.edu/abs/1966Natur.211..695S}{211, 695}

\bibitem[{{Sturrock}(1968)}]{1968Sturrock}
{Sturrock}, P.~A. 1968, in Structure and Development of Solar Active Regions,
  ed. K.~O. {Kiepenheuer}, Vol.~35,
  \href{https://ui.adsabs.harvard.edu/abs/1968IAUS...35..471S}{471}

\bibitem[{{Sweet}(1958{\natexlab{a}})}]{1958SweetNeutralPoint}
{Sweet}, P.~A. 1958{\natexlab{a}}, in Electromagnetic Phenomena in Cosmical
  Physics, ed. B.~{Lehnert}, Vol.~6,
  \href{https://ui.adsabs.harvard.edu/abs/1958IAUS....6..123S}{123}

\bibitem[{{Sweet}(1958{\natexlab{b}})}]{1958SweetParticles}
{Sweet}, P.~A. 1958{\natexlab{b}},
  \href{http://dx.doi.org/10.1007/BF02962520}{\color{magenta}Il Nuovo Cimento},
  \href{https://ui.adsabs.harvard.edu/abs/1958NCim....8S.188S}{8, 188}

\bibitem[{{Syrovatskii} \& {Shmeleva}(1972)}]{1972Syrovatskii}
{Syrovatskii}, S.~I. \& {Shmeleva}, O.~P. 1972, \sovast,
  \href{https://ui.adsabs.harvard.edu/abs/1972SvA....16..273S}{16, 273}

\bibitem[{{van Noort} {et~al.}(2005){van Noort}, {Rouppe van der Voort}, \&
  {L{\"o}fdahl}}]{vanNoort05}
{van Noort}, M., {Rouppe van der Voort}, L., \& {L{\"o}fdahl}, M.~G. 2005,
  \href{http://dx.doi.org/10.1007/s11207-005-5782-z}{\color{magenta}\solphys},
  \href{https://ui.adsabs.harvard.edu/abs/2005SoPh..228..191V}{228, 191}

\bibitem[{{Vernazza} {et~al.}(1976){Vernazza}, {Avrett}, \&
  {Loeser}}]{vernazza76}
{Vernazza}, J.~E., {Avrett}, E.~H., \& {Loeser}, R. 1976,
  \href{http://dx.doi.org/10.1086/190356}{\color{magenta}\apjs},
  \href{https://ui.adsabs.harvard.edu/abs/1976ApJS...30....1V}{30, 1}

\bibitem[{{Vissers} \& {Rouppe van der Voort}(2012)}]{Gregal12}
{Vissers}, G. \& {Rouppe van der Voort}, L. 2012,
  \href{http://dx.doi.org/10.1088/0004-637X/750/1/22}{\color{magenta}\apj},
  \href{https://ui.adsabs.harvard.edu/abs/2012ApJ...750...22V}{750, 22}

\bibitem[{{Vissers} {et~al.}(2021){Vissers}, {Danilovic}, {de la Cruz
  Rodr{\'\i}guez}, {Leenaarts}, {Morosin}, {D{\'\i}az Baso}, {Reid}, {Pomoell},
  {Price}, \& {Inoue}}]{2021gragal}
{Vissers}, G.~J.~M., {Danilovic}, S., {de la Cruz Rodr{\'\i}guez}, J., {et~al.}
  2021,
  \href{http://dx.doi.org/10.1051/0004-6361/202038900}{\color{magenta}\aap},
  \href{https://ui.adsabs.harvard.edu/abs/2021A&A...645A...1V}{645, A1}

\bibitem[{{{\v{S}}vestka} {et~al.}(1961){{\v{S}}vestka}, {Kopeck{\'y}}, \&
  {Blaha}}]{1961Svestka_flare_asymmetries}
{{\v{S}}vestka}, Z., {Kopeck{\'y}}, M., \& {Blaha}, M. 1961, Bulletin of the
  Astronomical Institutes of Czechoslovakia,
  \href{https://ui.adsabs.harvard.edu/abs/1961BAICz..12..229S}{12, 229}

\bibitem[{{{\v{S}}vestka} {et~al.}(1962){{\v{S}}vestka}, {Kopeck{\'y}}, \&
  {Blaha}}]{1962Svestka_flare_asymmetries}
{{\v{S}}vestka}, Z., {Kopeck{\'y}}, M., \& {Blaha}, M. 1962, Bulletin of the
  Astronomical Institutes of Czechoslovakia,
  \href{https://ui.adsabs.harvard.edu/abs/1962BAICz..13...37S}{13, 37}

\bibitem[{Wilson(1774)}]{wilson1774observations}
Wilson, A. 1774, Philosophical Transactions of the Royal Society of London

\bibitem[{{Wilson}(1969)}]{1969Wilson}
{Wilson}, P.~R. 1969,
  \href{http://dx.doi.org/10.1007/BF00145527}{\color{magenta}\solphys},
  \href{https://ui.adsabs.harvard.edu/abs/1969SoPh...10..404W}{10, 404}

\bibitem[{{Wuelser} \& {Marti}(1989)}]{wuelser1989}
{Wuelser}, J.-P. \& {Marti}, H. 1989,
  \href{http://dx.doi.org/10.1086/167567}{\color{magenta}\apj},
  \href{https://ui.adsabs.harvard.edu/abs/1989ApJ...341.1088W}{341, 1088}

\bibitem[{{Xu} {et~al.}(2023){Xu}, {Kerr}, {Polito}, {Huang}, {Jing}, \&
  {Wang}}]{Xu2023}
{Xu}, Y., {Kerr}, G.~S., {Polito}, V., {et~al.} 2023,
  \href{https://ui.adsabs.harvard.edu/abs/2023arXiv230905745X}{\href{http://dx.doi.org/10.48550/arXiv.2309.05745}{\color{magenta}arXiv
  e-prints}, arXiv:2309.05745}

\bibitem[{{Yadav} {et~al.}(2021){Yadav}, {D{\'\i}az Baso}, {de la Cruz
  Rodr{\'\i}guez}, {Calvo}, \& {Morosin}}]{Yadav21}
{Yadav}, R., {D{\'\i}az Baso}, C.~J., {de la Cruz Rodr{\'\i}guez}, J., {Calvo},
  F., \& {Morosin}, R. 2021,
  \href{http://dx.doi.org/10.1051/0004-6361/202039857}{\color{magenta}\aap},
  \href{https://ui.adsabs.harvard.edu/abs/2021A&A...649A.106Y}{649, A106}

\bibitem[{{Zarro} \& {Canfield}(1989)}]{1989Zarro}
{Zarro}, D.~M. \& {Canfield}, R.~C. 1989,
  \href{http://dx.doi.org/10.1086/185394}{\color{magenta}\apjl},
  \href{https://ui.adsabs.harvard.edu/abs/1989ApJ...338L..33Z}{338, L33}

\bibitem[{{Zarro} {et~al.}(1988){Zarro}, {Canfield}, {Strong}, \&
  {Metcalf}}]{zarro1988}
{Zarro}, D.~M., {Canfield}, R.~C., {Strong}, K.~T., \& {Metcalf}, T.~R. 1988,
  \href{http://dx.doi.org/10.1086/165919}{\color{magenta}\apj},
  \href{https://ui.adsabs.harvard.edu/abs/1988ApJ...324..582Z}{324, 582}

\bibitem[{{Zharkov} {et~al.}(2020){Zharkov}, {Matthews}, {Zharkova}, {Druett},
  {Inoue}, {Dammasch}, \& {Macrae}}]{2020Zharkov}
{Zharkov}, S., {Matthews}, S., {Zharkova}, V., {et~al.} 2020,
  \href{http://dx.doi.org/10.1051/0004-6361/201936755}{\color{magenta}\aap},
  \href{https://ui.adsabs.harvard.edu/abs/2020A&A...639A..78Z}{639, A78}

\bibitem[{{Zharkova} {et~al.}(2011){Zharkova}, {Arzner}, {Benz}, {Browning},
  {Dauphin}, {Emslie}, {Fletcher}, {Kontar}, {Mann}, {Onofri}, {Petrosian},
  {Turkmani}, {Vilmer}, \& {Vlahos}}]{2011Zharkova}
{Zharkova}, V.~V., {Arzner}, K., {Benz}, A.~O., {et~al.} 2011,
  \href{http://dx.doi.org/10.1007/s11214-011-9803-y}{\color{magenta}\ssr},
  \href{https://ui.adsabs.harvard.edu/abs/2011SSRv..159..357Z}{159, 357}

\bibitem[{{Zharkova} \& {Kobylinskii}(1993)}]{1993Zharkova}
{Zharkova}, V.~V. \& {Kobylinskii}, V.~A. 1993,
  \href{http://dx.doi.org/10.1007/BF00646487}{\color{magenta}\solphys},
  \href{https://ui.adsabs.harvard.edu/abs/1993SoPh..143..259Z}{143, 259}

\bibitem[{{Zhu} {et~al.}(2019){Zhu}, {Kowalski}, {Tian}, {Uitenbroek},
  {Carlsson}, \& {Allred}}]{2019ZhuMgkhFlareRADYN}
{Zhu}, Y., {Kowalski}, A.~F., {Tian}, H., {et~al.} 2019,
  \href{http://dx.doi.org/10.3847/1538-4357/ab2238}{\color{magenta}\apj},
  \href{https://ui.adsabs.harvard.edu/abs/2019ApJ...879...19Z}{879, 19}

\end{thebibliography}

\end{document}